%% file: AZhPaper.tex
\documentclass[3p,twocolumn,sort&compress]{elsarticle}
\pdfoutput=1 
\DeclareGraphicsExtensions{.pdf}

\def\paperabstract{
A search  for a heavy, 
CP-odd Higgs boson, $A$,  decaying into a $Z$ boson and 
a 125~\gev\ Higgs boson, $h$, with the ATLAS detector at the LHC is presented. 
The search uses proton--proton collision data  
at a centre-of-mass energy of 8~TeV corresponding to an integrated
luminosity of 20.3~fb$^{-1}$. 
Decays of CP-even $h$ bosons to $\tau\tau$ or $bb$  pairs 
with the $Z$ boson decaying to electron or muon pairs are considered,  
as well as $h \rightarrow bb$ decays with the $Z$ boson decaying 
to neutrinos.  No evidence for the production of an $A$ boson 
in these channels is found and the 95\% confidence level upper limits derived 
for $\sigma (gg\to A) \times \text{BR}(A\to Zh) 
\times \text{BR}(h\to f\bar{f})$ are 0.098--0.013~pb for $f=\tau$ and  0.57--0.014~pb for $f=b$  
in a range of $m_A =$ 220--1000~GeV. 
The results are combined and interpreted in the context of two-Higgs-doublet models.
}

\usepackage[labelfont=bf]{caption}
\usepackage{preprintcover}
\usepackage{atlasphysics}
\usepackage{subfigure}
\usepackage{multirow}
\usepackage{amsmath}
\usepackage{amssymb}
\usepackage{lineno}
\usepackage{ifpdf}
\usepackage[hyperindex,breaklinks]{hyperref} 
\usepackage{verbatim}
\ifpdf\usepackage{grffile}\else\fi

\def\invfb {\ensuremath{{\mathrm{fb}^{-1}}}}

\def\met {\ensuremath{E^\mathrm{miss}_{\mathrm{T}}}}
\def\vecmet {\ensuremath{\vec{E}^{\mathrm{miss}}_{\mathrm{T}}}}
\def\vecmpt {\ensuremath{\vec{p}_{\mathrm{T}}^{\mathrm{\ miss}}}}

\def\mta {\ensuremath{m^{\text{rec},\text{T}}_{A}}}
\def\mpt {\ensuremath{{p_{\mathrm{T}}^{\mathrm{miss}}}}}
\def\to {\ensuremath{\rightarrow}}

\def\lephad {\ensuremath{\tau_{\mathrm{lep}}\tau_{\mathrm{had}}}}
\def\hadhad {\ensuremath{\tau_{\mathrm{had}}\tau_{\mathrm{had}}}}
\def\leplep {\ensuremath{\tau_{\mathrm{lep}}\tau_{\mathrm{lep}}}}
\def\thad {\ensuremath{\tau_{\mathrm{had}}}}
\def\tlep {\ensuremath{\tau_{\mathrm{lep}}}}

\def\ZZstar{\ensuremath{ZZ^{(\ast)}}}

\def\ptZ{\ensuremath{p^{Z}_{\mathrm{T}}}}

\def\ETMISS{\ensuremath{E^\mathrm{miss}_{\mathrm{T}}}}
\def\pt{\ensuremath{p_{\mathrm{T}}}}
\def\Et{\ensuremath{E_{\mathrm{T}}}}

\newif\ifnotpagetest
\notpagetesttrue

\ifnotpagetest
\else
\fi

\def\papertitle{Search for a CP-odd Higgs boson decaying to $Zh$ in $pp$ 
collisions at $\sqrt{s} = 8$~TeV with the ATLAS detector}

\PreprintCoverPaperTitle{\papertitle}
\PreprintIdNumber{CERN-PH-EP-2015-013} 
\PreprintJournalName{Phys. Lett. B}
\PreprintCoverAbstract{\paperabstract}

\journal{Journal} 

\begin{document}

\begin{frontmatter}
\title{\papertitle}
\author{The ATLAS Collaboration}
\address{}
\begin{abstract} 
\ifnotpagetest
\paperabstract
\fi
\end{abstract}
\end{frontmatter}

\section{Introduction}
\label{sec:intro}
\input{introduction}

\section{Data and simulated samples}
\label{sec:samples}
\input{samples}

\section{Object reconstruction}
\label{sec:det}
\input{detector}

\section{Search for $A\to Zh$ with $h \rightarrow \tau\tau$}
\label{sec:lltautau}
\input{lltautau}

\subsection{$\ell\ell\hadhad$}
\label{sec:llhadhad}
\input{llhadhad}

\subsection{$\ell\ell\lephad$}
\label{sec:lllephad}
\input{lllephad}

\subsection{$\ell\ell\leplep$}
\label{sec:llleplep}
\input{llleplep}

\subsection{Systematic uncertainties and results}
\label{sec:lltautau_summary}
\input{lltautauSummary}

\section{Search for $A\to Zh$  with $h \rightarrow bb$}
\label{sec:xxbb}
\input{xxbb}

\subsection{$\ell \ell b b$ selection}
\label{sec:llbb}
\input{llbb}

\subsection{$\nu \nu b b$ selection}
\label{sec:nnbb}
\input{nnbb}

\subsection{Backgrounds}
\label{sec:bbbackgrounds}
\input{bbbackgrounds}

\subsection{Systematic uncertainties and results}
\label{sec:bbsummary}
\input{bbsummary}

\input{results}

\section{Conclusions}
\label{sec:summary}
\input{summary}

\input{Acknowledgements}

\section*{References}
\bibliographystyle{atlasBibStyleWoTitle}
\bibliography{AZhPaper}

\onecolumn 
\clearpage
\input{atlas_authlist}

\end{document}
\bye

%% file: introduction.tex
After the discovery of a Higgs boson at the LHC in
2012~\cite{paper2012ichep,CMSpaper}, one of the most important remaining 
questions is whether the newly discovered particle is part of an
extended scalar sector. A CP-odd Higgs boson, $A$, appears in many
models with an extended scalar sector, e.g. in the case of the 
two-Higgs-doublet model (2HDM)~\cite{Branco:2011iw}. 

The addition of a second Higgs doublet leads to five Higgs bosons after the
electroweak symmetry breaking. The phenomenology of such a model is very rich
and depends on the vacuum expectation values of the Higgs doublets,
the CP properties of the Higgs potential and the values of its parameters
and the Yukawa couplings of the Higgs doublets with the fermions.
In general, it is possible to accommodate in the model a Higgs boson compatible
to the one discovered at the LHC.
In the case where the Higgs potential of the 2HDM is 
CP-conserving, the Higgs bosons after electroweak symmetry breaking are
 two CP-even ($h$ and $H$), one CP-odd ($A$) and
two charged ($H^{\pm}$) Higgs bosons. 
Many theories beyond the Standard Model (SM)  include a second Higgs doublet, 
such as the minimal supersymmetric SM (MSSM)
\cite{Fayet:1976et,Fayet:1977yc,Farrar:1978xj,Fayet:1979sa,Dimopoulos:1981zb},
axion models (e.g. Ref. \cite{Kim:1986ax}) and baryogenesis models 
(e.g. Ref. \cite{Joyce:1994zt}).
Searches for a CP-odd Higgs boson are reported 
in Refs.~\cite{Schael:2006cr,Aaltonen:2012zh,Aad:2014vgg,Khachatryan:2014wca}.

In this Letter, a search for a heavy CP-odd Higgs boson decaying into a
$Z$ boson and the  $\sim$125 GeV Higgs boson, $h$, is described. 
The $A\to Zh$ decay rate can be dominant for part of the 2HDM parameter 
space, especially for an $A$ boson mass, $m_A$,  below the $t\bar{t}$ threshold.
In this case, the $A$ boson is produced mainly via gluon fusion
and its natural width is typically small: 
$\Gamma_A/m_A \lesssim \mathcal{O}(1\%)$.

The search is performed for $m_A$ in the range 220 to 1000~\gev, 
reconstructing\footnote{Throughout this Letter, the notation $h\rightarrow bb$,
 $h\rightarrow \tau\tau$, $Z\rightarrow \nu\nu$ and $Z\rightarrow \ell\ell$ 
is used for $h\rightarrow b\bar{b}$, $h\rightarrow \tau^{+}\tau^{-}$, 
$Z\rightarrow \nu\bar{\nu}$ and $Z\rightarrow \ell^{+}\ell^{-}$, respectively.}
$Z \to \ell \ell$ decays (where $\ell=e, \mu$)  with $ h\to bb$ 
or $h\to \tau\tau$, as well as $ Z \to \nu\nu$ with $h\to bb$.
The selected $h$ boson decay modes provide high branching ratios 
and the possibility to fully reconstruct the Higgs boson decay kinematics.
The reconstructed invariant mass (or transverse mass) of the $Zh$ pair, 
employing the measured value of the $h$ boson mass, $m_h$, to improve its resolution, 
is used to search for a signal.

%% file: samples.tex
The data used in this search were recorded with the
ATLAS detector in proton--proton collisions at a centre-of-mass
energy of 8 TeV.
The ATLAS detector is described in detail elsewhere~\cite{Aad:2008zzm}. 
The integrated luminosity of the data sample, selecting only
periods where all relevant detector subsystems were operational,
is  20.3~$\pm$~0.6~$\invfb$~\cite{Aad:2013ucp}.
The data used in the $\ell\ell\tau\tau$ and $\ell\ell bb$ final states
were collected using a combination of single-electron, single-muon, dielectron ($ee$) and dimuon ($\mu\mu$) triggers. 
Depending on the trigger choice, the $\pt$\footnote{ATLAS uses a right-handed coordinate 
system with its origin at the nominal interaction point (IP) in the centre of the detector 
and the $z$-axis along the beam pipe. The $x$-axis points from the IP to the centre of the LHC ring, 
and the $y$-axis points upward. Cylindrical coordinates $(r,\phi)$ are used in the transverse 
plane, $\phi$ being the azimuthal angle around the beam pipe. The pseudorapidity is defined in 
terms of the polar angle $\theta$ as $\eta=-\ln\tan(\theta/2)$. Transverse momenta are computed from
the three-momenta, $\vec{p}$, as $\pt = |\vec{p}| \sin\theta $.} thresholds
vary from 24 to 60~GeV for the single-electron and single-muon triggers,  
and from 12 to 13~GeV
for the $ee$ and $\mu\mu$ triggers.
The data used in the $\nu\nu bb$ final state were collected
with a missing transverse momentum ($\ETMISS$) trigger 
with a threshold of $\ETMISS > 80$~GeV.

Signal events from a narrow-width $A$ boson produced
via gluon fusion are generated with {\tt MadGraph5} \cite{Alwall:2011uj}
for all final states considered in this search. 
The parton showering is performed with 
{\tt PYTHIA8} \cite{Pythia,Pythia8}.

Production of $W$ and $Z$ bosons in association with jets is simulated
with {\tt SHERPA}~\cite{SHERPA}. Top-quark pair and single top-quark production 
is simulated with {\tt POWHEG}~\cite{Powheg1,Powheg2,Powheg3} 
and {\tt AcerMC}~\cite{AcerMC}.
Production of $WW$, $WZ$, and $ZZ$ dibosons are simulated using {\tt POWHEG}. 
The $WZ$ and $ZZ$ processes include the production of off-shell $Z$ bosons 
($Z^*$) and photons ($\gamma^{*}$).
Triboson production ($WWW^{(*)}$, $ZWW^{(*)}$, $ZZZ^{(*)}$)  
and top pair production in association with a $Z$ boson
are generated with {\tt MadGraph5}.
Finally, the production of the SM Higgs boson in association with a $Z$ 
boson is considered as a background in this search. It is simulated using {\tt PYTHIA8}.

The {\tt CTEQ6L1} \cite{Pumplin:2002vw} set of parton distribution functions 
was used for samples generated with {\tt MadGraph5} and {PYTHIA8}.
The {\tt CT10} \cite{Lai:2010vv} set was used for the other samples.

All generated samples are passed through the GEANT4-based \cite{AGO-0301} 
detector simulation of the ATLAS detector~\cite{ATLASSIM}.
The simulated events are  overlaid with  minimum-bias events,
to account for the effect of multiple interactions occurring in the same 
and neighbouring bunch crossings (``pile-up''). The events
are reweighted so that the average number of interactions
per bunch crossing agrees with the data.

The background estimation in this search for most processes is based 
on data driven techniques, but in some cases only simulated samples are used. 
In that case, the simulated samples are normalized using theoretical cross 
section calculations. In particular, for diboson production both
$q\bar{q}$~\cite{Campbell:1999ah} and $gg$~\cite{Binoth:2006mf,Binoth:2008pr} 
initiated processes are included. 
Triboson production follows Ref.~\cite{Binoth:2008kt} 
and top pair production in association with a $Z$ boson
follows Refs.~\cite{Campbell:2012dh, Garzelli:2012bn}. 
SM Higgs boson production in association with a $Z$ boson uses a 
calculation described in Ref.~\cite{Heinemeyer:2013tqa}.

%% file: detector.tex
Electrons are identified from energy clusters in the electromagnetic 
calorimeter that are matched to tracks in the inner detector 
\cite{Aad:2014fxa}.
 Electrons are required to have $|\eta| < 2.47$
and $\pt > 7$~GeV. Isolation requirements, defined 
in terms of the calorimetric energy or the $\pt$ of tracks within cones
around the object, as well as quality requirements are  
applied to distinguish electrons from jets. 

Muons are reconstructed by matching tracks reconstructed in 
the inner detector to tracks or track segments in the muon spectrometer systems \cite{Aad:2014rra}.
The muon acceptance is extended to the region $ 2.5 < |\eta| < 2.7$, which 
is outside the inner detector coverage, using only tracks reconstructed in
the forward part of the muon detector. Muons used for this search must have 
$|\eta| < 2.7$, $\pt > 6$~GeV and are also required to pass isolation requirements.

Jets are reconstructed using the anti-$k_t$
algorithm~\cite{Cacciari:2008gp} with radius parameter \mbox{$R = 0.4$}
and $\pt>20$~GeV ($\pt>30$~GeV) for $|\eta| < 2.5$ ($2.5< |\eta| <
4.5$). Low-$\pt$ jets from pile-up are rejected with a requirement on
the scalar sum of the $\pt$ of the tracks associated with the jet: for
jets with $|\eta| < 2.4$ and $\pt < 50$~GeV, tracks associated with
the primary vertex\footnote{The primary vertex is taken to be the
reconstructed vertex with the highest $\Sigma\pt^{2}$ of the associated
tracks.} 
must contribute over 50\% to the sum.  Jets from
the decay of long-lived heavy-flavour hadrons are selected using a
multivariate tagging algorithm
($b$-tagging)~\cite{ATLAS-CONF-2014-004}.  The $b$-tagging efficiency
is 70\% for jets from $b$-quarks in a sample of simulated $t\bar{t}$ events.

Hadronic decays of $\tau$ leptons (\thad)~\cite{taus} 
are reconstructed starting from clusters of energy in the calorimeter.
A \thad{} candidate must lie within $|\eta| < 2.47$,  have a
transverse momentum greater than 20~GeV, one or three associated tracks
and a total charge of $\pm 1$. Information on the collimation, isolation,
and shower profile is combined into a multivariate discriminant
 to reduce backgrounds from quark- or gluon-initiated jets. 
Dedicated algorithms that reduce the number of electrons and muons
misidentified as hadronic $\tau$ decays are applied.
In this analysis, two \thad{} identification selections are used
---``loose'' and ``medium''--- with efficiencies
of about 65\% and 55\%, respectively.

The missing transverse momentum ($\vecmet$) is computed using fully calibrated
and reconstructed physics objects, as well as clusters of calorimeter-cell energy deposits 
that are not associated with any object~\cite{ATLAS-CONF-2013-082}. 
In addition, a track-based missing transverse momentum ($\vecmpt$)
is calculated as the negative vector sum of the transverse momenta of
tracks with  $|\eta| < 2.4$ and associated with the primary vertex.

%% file: lltautau.tex
\label{sec:lltautau_common}

In  the search for $A \to Zh \to \ell\ell \tau\tau$, 
three channels are considered, distinguished by the way
the $\tau\tau$ pair decays: two $\tau$ leptons decaying 
hadronically ($\thad\thad$),
one  leptonic and one hadronic decay 
($\tlep\thad$) and, finally, two leptonic decays
($\tlep\tlep$).
Electrons in the  $\thad\thad$ and $\tlep\thad$
channels are rejected in the 
transition region between the barrel and end-cap of the detector 
($1.37 < |\eta| < 1.52$). Muons in the $\thad\thad$ and $\tlep\thad$
channels are considered only for $|\eta| < 2.5$.

The resolution of the reconstructed $A$ boson mass is improved using
a  mass-difference variable, 
\begin{eqnarray*}
m_{A}^{\text{rec}} = m_{\ell\ell\tau\tau} - m_{\ell\ell} -m_{\tau\tau} +m_{Z} + m_{h},
\end{eqnarray*}
where $m_{Z}$ is the mass of the $Z$ boson,
$m_{h} = 125$~GeV is the mass of the CP-even Higgs boson, $m_{\ell\ell}$ is 
the invariant mass of the two leptons associated with the $Z$ boson decay, 
and $m_{\ell\ell\tau\tau}$ denotes the $\ell\ell\tau\tau$ invariant
mass. The value of $m_{\tau\tau}$, the invariant mass of the $\tau$'s, 
is estimated with the Missing Mass Calculator (MMC) \cite{MMC}. 
The mass resolution for all $\tau\tau$ channels ranges from 3\% at $m_A=220$~\gev\ to 5\% at $m_A=1$~\tev.

%% file: llhadhad.tex
Events in  the $\ell\ell\hadhad$ channel
are required to contain exactly two  opposite-sign leptons 
$\ell \ell$ ($ee$ or $\mu\mu$)
and exactly two opposite-sign $\thad$. The $\pt$ requirements for these objects
are $\pt > 26$~GeV (15~GeV) for the leading (subleading) electron,
 $\pt > 25$--$36$~GeV (10~GeV) for the leading (subleading) muon, depending
on the trigger, and $\pt > 35$~GeV (20~GeV) for the leading 
(subleading) $\thad$ candidates. 
The $\thad$ candidates are required to satisfy the ``loose'' $\thad$ identification
criterion.
In addition, the $ee/\mu\mu$ invariant mass 
and the $\tau\tau$ invariant mass have to lie in the ranges
 $80 < m_{\ell\ell} < 100$~GeV and $75<m_{\tau\tau}<175$~GeV.
Finally, the $\pt$ of the $\ell\ell$ pair, \ptZ, is required to be:
\begin{eqnarray*}
\ptZ > \left\{
  \begin{array}{l}
    125~\text{GeV}, \ \text{if}\  m_{A}^{\text{rec}} > 400~\text{GeV} \\
    0.64 \times m_{A}^{\text{rec}} - 131~\text{GeV},\ \text{otherwise}.

  \end{array} \right.
\end{eqnarray*}
This requirement  maximizes the 
sensitivity over the whole explored $A$ mass range. 
In the region of $\ptZ >$ 125 GeV, there is little background present, 
so tightening the requirement results in no additional increase in sensitivity. The total acceptance times selection 
efficiency varies from 6.2\%, for $m_A = 220$~\gev, to around 18\% for the highest $A$ boson masses considered.

The dominant background for this channel originates from events where one 
or both of the $\thad$'s is a misidentified jet (``fake-$\thad$ background'').
This background is dominated by $Z$+jets events, 
with small contributions from dibosons and  events with top quarks,
and it is estimated using a template method.
The shape of the fake-$\thad$ background is taken from a control region
(the ``template region'') that contains events satisfying 
all the $\ell\ell\thad\thad$  selection criteria apart from the 
requirements for an opposite-sign  $\thad\thad$ pair and the 
$\thad$ identification criteria.
The fake-$\thad$ background is normalized by using two additional
control regions. The first region, ``A'', contains events
that satisfy the {\it signal} selection criteria, with the exception
that the $m_{\tau\tau}$ constraint is inverted, i.e.
$m_{\tau\tau} < 75$~GeV or $m_{\tau\tau} > 175$~GeV.
The second region, ``B'', contains events that satisfy all the
{\it template}  selection criteria, with the exception
that the $m_{\tau\tau}$ constraint is inverted, as in the 
region  ``A'' definition.
The ratio of the number of events in ``A'' to 
the number of events in ``B'' is used to scale the template
region events in order to obtain the normalization of the 
fake-$\thad$ background.

In addition to the fake-$\thad$ background, 
there are also contributions from backgrounds with
real $\ell\ell\thad\thad$ objects in the event.
These backgrounds come primarily from $\ZZstar$ production\footnote{The notation
$\ZZstar$ is used here to include $ZZ$, $ZZ^{\star}$ and $Z\gamma^{\star}$}. 
SM Higgs boson production in association with a Z boson is 
estimated using simulation, and contributes 17\% of 
the total background.

%% file: lllephad.tex
Events in the $\ell\ell\tlep\thad$ channel are required to
contain exactly three light leptons, $\mu\mu\mu$, $e\mu\mu$, $ee\mu$ or $eee$,
and exactly one $\thad$. The $\pt$ requirements for these objects
are $\pt > 26$~GeV (15~GeV) for the leading (remaining) electron(s),
 $\pt > 25$--36~GeV (10~GeV) for the leading (remaining) muon(s), depending
on the trigger, and $\pt > 20$~GeV for the $\thad$. 
Subsequently, all the possible $\ell\ell$ pairs
that are composed of opposite-sign, same-flavour leptons
are selected. From these pairs, the pair that has the 
invariant mass closest to  $m_Z$ 
is considered to be the lepton pair from the $Z$ boson decay.
The third light lepton is considered to be the leptonic $\tau$ decay,
and it is used along with the $\thad$ to define the $\tlep\thad$ pair.
This light lepton is required to have opposite-sign charge with respect to the $\thad$.
In addition, the $\thad$ is required to satisfy the ``medium'' $\thad$
identification requirement, and  $m_{\ell\ell}$ 
and $m_{\tau\tau}$  have to lie in the ranges
 $80 < m_{\ell\ell} < 100$~GeV and $75<m_{\tau\tau}<175$~GeV. 
The total acceptance times selection efficiency varies from 6\% for $m_A = 220$~\gev, to
around 17\% for the highest $A$ boson masses considered.

About half of the total background for this channel comes from events
where the $\thad$ and/or the light lepton is a misidentified jet
(``fake-$\tau/\ell$ background''). This background is dominated by
diboson and $Z$+jets events and it is estimated using a template method.
The shape of the fake-$\tau/\ell$ background is taken from a control region
(the ``template region'') that contains events satisfying 
all $\ell\ell\tlep\thad$ selection criteria,
apart from requiring ``medium'' $\thad$ identification criterion and 
opposite-sign charge for the $\tlep\thad$ pair. 
The fake-$\tau/\ell$ background is normalized by using two additional
control regions, defined similarly to those in the
$\ell\ell\thad\thad$ channel.

The other half of the background comes from events
with real $\ell\ell\tlep\thad$ objects in the
event. These backgrounds come primarily from
$\ZZstar$ production. 
There is also a small (11\%) contribution from the SM Higgs boson production
in association with a $Z$ boson, which is estimated using simulation.

%% file: llleplep.tex
Events in the $\ell\ell\tlep\tlep$ channel are required to
contain at least four leptons, which form 
one same-flavour and opposite-sign  pair consistent with the
$Z$ mass ($80 < m_{\ell\ell} < 100$~\GeV), 
and either a same-flavour or different-flavour 
pair with an invariant mass reconstructed with the MMC algorithm,
consistent with a decay from the CP-even Higgs boson (
$90 < m_{\tau\tau} <190\, \gev$).
One muon is allowed to be reconstructed in the forward region 
($2.5<|\eta|<2.7$) of the muon spectrometer,
or to be identified in the calorimeter with $\pt>15$\,\gev\ and  
$|\eta|<0.1$~\cite{Aad:2014rra}.
The highest-\pt~lepton 
must satisfy $\pt >20$\,\gev, and the second (third) lepton in $\pt$
order must satisfy $\pt >15$\,\gev\ ($\pt >10$\,\gev). 
Among all the possible lepton quadruplets in an event 
the one minimizing the sum of the mass differences with respect to  both the $Z$ and $h$ bosons is chosen.

Two different analysis categories are defined based on the lepton flavours in the Higgs 
boson decay: $ee$ or $\mu\mu$ (SF), and $e\mu$ (DF).
The expected background is very different in the two cases.
For the SF channel, the background is dominated by \ZZstar\ production with $Z\to ee/\mu\mu$ decays. 
For the DF channel, the main background is from the \ZZstar\ process 
through the $Z\rightarrow\tlep\tlep$ decay chain, but other backgrounds are also important.
 The signal-to-noise ratio in the SF category is improved by using a set of requirements specifically targeted to suppress  the main \ZZstar\ background.  First, 
a veto on the on-shell production of $Z$ boson pairs is introduced, requiring the invariant mass of the $h$ boson leptons to lie outside the 
$Z$ peak: $m_h < 80 \gev$ or $m_h > 100 \gev$. 
Background events are characterized by low missing transverse momentum and are further rejected by requiring $\met > 30\,\gev$, and  
the azimuthal angle between  
the \met{} direction and the $Z$ boson transverse momentum to be greater than $\pi/2$. 
Furthermore, a requirement that the highest-\pt{}
lepton of the $\ell\ell$ pair associated with the 
 $h$ boson has $\pt > 15$~GeV is applied, since it is found to be
effective against backgrounds from
$Z$+jets production. The total acceptance times selection efficiency varies from 6.5\% (1.5\%) for DF (SF) channel for $m_A = 220$~\gev, to 
around 20\% for both channels for the highest $A$ boson masses considered.

The subleading contributions to the background are from 
diboson and triboson production, $t\bar{t}$ production in association 
with a $Z$ boson, and SM Higgs boson production. All these
are determined from simulation and amount 
to about 95\% (65\%) of the total background in the SF (DF) category. 
The other background events have at least one lepton which is a
misidentified jet or a lepton from a heavy-flavour
quark decay and are dominated by $Z$+jets production,
with a smaller contribution from top-quark production.
These backgrounds are estimated using a control
region  where one or both of the leptons in the
$\ell\ell$ pair associated with the $h\rightarrow \tlep\tlep$ 
decay fail to satisfy the
isolation criteria. After subtraction of genuine sources 
of four-lepton events using simulation, the data
are extrapolated to the \emph{isolated} signal region
using normalization factors derived from simulated samples.

%% file: lltautauSummary.tex
\begin{figure*} 
  \centering
  \subfigure[$\ell\ell\lephad$ + $\ell\ell\hadhad$]{
    \includegraphics[width=0.48\textwidth]{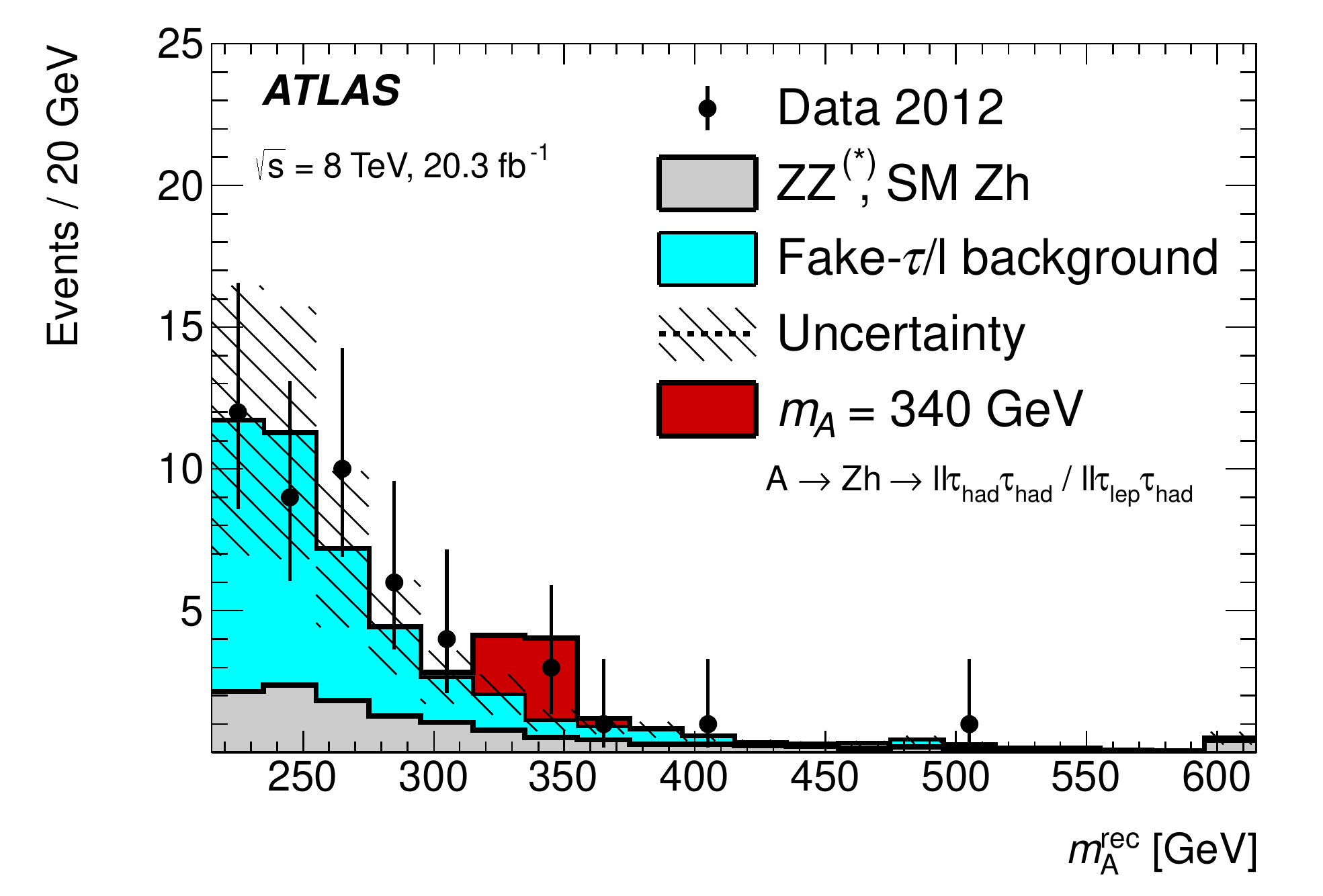}
  }
  \subfigure[$\ell\ell\leplep$]{
    \includegraphics[width=0.48\textwidth]{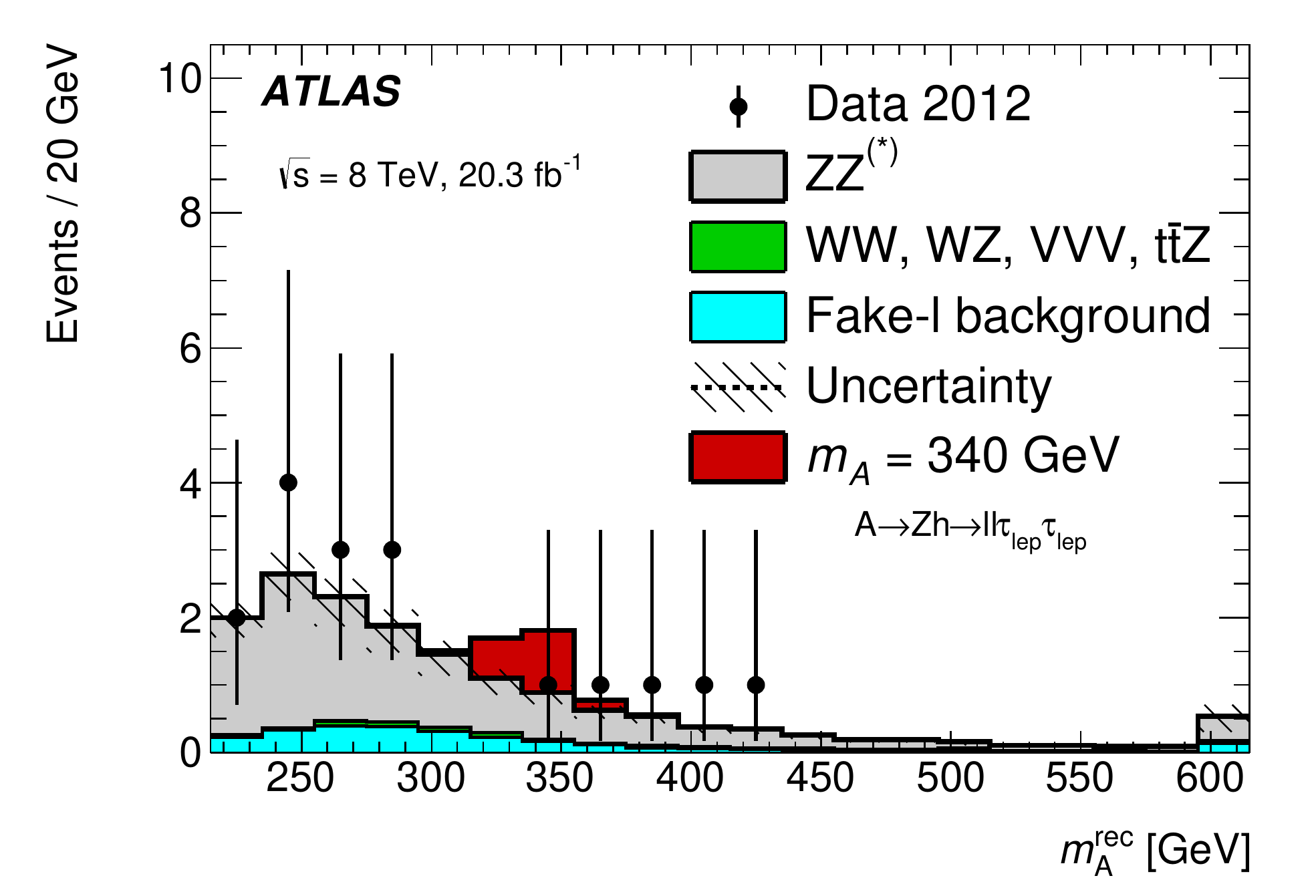}
  }

  \caption{Distributions of the reconstructed $A$ boson mass 
for the combined $\ell\ell \hadhad$ and $\ell\ell \lephad$ final states (a) 
and the $\ell\ell\leplep$ final states (b). The signal shown in both cases corresponds to 
$\sigma (gg\to A) \times \text{BR}(A\to Zh) \times \text{BR}(h\to \tau\tau) = 50$~fb 
with $m_A$ = 340~GeV. The background contributions shown are the results of simulation and  data-driven estimation
methods. The background uncertainty is shown as a hatched area, and the overflow is included in the last bin.
}
  \label{fig:tautau_final}
\end{figure*}

The most important systematic uncertainty for the backgrounds with real 
$\ell\ell\tau\tau$ objects in the
$\tlep\thad$ and $\tlep\tlep$ channels comes from the uncertainty on the 
theoretical cross sections used in the normalization. They are 
due to the parton distribution function choice, the renormalization 
and factorization scales, as well as
the \alphas\ value. This amounts to an uncertainty on the normalization of this 
background of about 5.0\% for the $\tlep\thad$ channel and 6.4\% for $\tlep\tlep$. 
In the $\thad\thad$ channel, the largest contributions come from the $\thad$ 
identification and energy scale and amounts to 8.9\%~\cite{taus}.
The fake-$\thad/\ell$ background systematic uncertainty for the
$\tau\tau$ channels is dominated by
the statistical uncertainty on data in control regions used for the
background normalization. It amounts to a normalization
uncertainty of 38\% and 25\% for the 
$\tlep\thad$ and $\thad\thad$ channels, respectively.
For the $\tlep\tlep$ channel, the normalization uncertainty
is 65\% (25\%) for the SF  (DF) category.

The reconstructed $A$ boson mass distributions 
for events passing the $\ell\ell\thad\thad$,
 $\ell\ell\tlep\thad$ and $\ell\ell\tlep\tlep$
selections are shown in Fig.~\ref{fig:tautau_final}.
The number of events passing the  $\ell\ell\tau\tau$ channel selections
are shown in Table~\ref{tab:lltautau_event_yields}. 
The agreement of the  expectation with data is very good.

\begin{table} [h!]
  \caption{The number of predicted and observed
events for the $\ell\ell\tau\tau$ channels. 
}
  \centering
  \begin{tabular}{lcc}

\hline
          &   Expected Background  & Data \\
\hline
$\ell\ell\thad\thad$ &   $28\pm 6$\phantom{0}  &   29 \\
$\ell\ell\tlep\thad$ &   $17\pm 4$\phantom{0}  &   18  \\
$\ell\ell\tlep\tlep$ (SF)&   $9.5\pm 0.6$        &  10 \\
$\ell\ell\tlep\tlep$ (DF)&   $7.2\pm 0.7$        &  7 \\

\hline
  \end{tabular}

  \label{tab:lltautau_event_yields}
\end{table}

%% file: xxbb.tex
This section describes the searches in the $A \rightarrow Zh \rightarrow
\ell \ell bb$ and $A \rightarrow Zh \rightarrow \nu\nu bb$ channels.

%% file: llbb.tex
Events in the $\ell\ell bb$ channel are selected by requiring either
two electrons or two muons. In the case of muons they are required to
be of opposite-sign charge. Leptons must have $\pt>7$~GeV, and electrons are
restricted to $|\eta|<2.47$, while muons must have $|\eta|<2.7$.
Tighter acceptance requirements are placed on one of the
leptons in each event in order to select a sample for which the trigger
efficiency is high and to reduce the multi-jet background, 
while keeping a high signal acceptance. These requirements are that
the leptons have $\pt>25$~GeV, and, if they are  muons, satisfy $|\eta|<2.5$.
A dilepton invariant mass window of $83<m_{\ell\ell}<99$~GeV 
is imposed to reduce top-quark and multi-jet backgrounds.

The $h \rightarrow bb$ decay is reconstructed by requiring two
$b$-tagged jets with $\pt>45$~GeV (20~GeV) for the leading
(subleading) jet. Events with more than two $b$-tagged jets are removed but all
events with one or more additional jets failing $b$-tagging are retained. 
The $h\rightarrow bb$ decay is selected by requiring that
the invariant mass of the two $b$-tagged jets lies within the range
 $105<m_{bb}<145$~GeV.

The top-quark background, which includes top-quark pair and single top-quark production, 
is reduced by requiring $\met/\sqrt{H_{\text{T}}}<3.5 \gev^{1/2}$,  where $H_{\text{T}}$ is
defined as the scalar sum of the $\pt$ of all jets and leptons in the event.

The reconstructed $A$ boson mass, $m_A^{\text{rec}}$, is 
the invariant mass of the two leptons and two $b$-tagged jets. In this 
calculation, the four-momentum of each $b$-tagged jet is scaled by $125~\text{GeV}/m_{bb}$ in order to
improve the resolution. 
The resulting $m_A^{\text{rec}}$  resolution  ranges from 2\% at $m_A=220$ \gev\ to 3\% at $m_A=1$ \tev.

In order to reduce the dominant $Z+$jets background, a requirement
is imposed on the transverse momentum of the $Z$ boson, $\ptZ$, 
reconstructed from the two leptons: 
$\ptZ >0.44 \times m_A^{\text{rec}}-106$~GeV, where
$m_A$ is in units of GeV. 
The requirement depends on $m_A^{\text{rec}}$ since the
background is generally produced at low $\ptZ$, 
whereas the mean
$\ptZ$ increases with $m_A$ for the signal. The total acceptance times selection
efficiency varies from 7\%, for $m_A = 220$~\gev, to around 16\% for the highest $A$ boson masses considered.

%% file: nnbb.tex
The event selection in the $\nu \nu bb$ channel follows
closely the SM $h\to bb$ analysis in Ref.~\cite{Aad:2014xzb}. 
Events are selected with $\met>120$~GeV, $\mpt>30$~GeV
 and no electrons or muons with $\pt>7$~GeV.  
In addition to the jet selection of the $\ell\ell bb$ 
analysis, additional restrictions are applied.
In order to suppress top-quark background, which is larger 
than in the $\ell\ell bb$ channel, events
are rejected if any of the following conditions is satisfied: 
there is a jet with $|\eta|>2.5$; there are
four or more jets; one  of the $b$-tagged jets is the third-highest-$\pt$ jet.  
In order to select a sample for which the trigger efficiency is high,
$\HT$ is required to be above $120$~GeV ($150$~GeV) for events with 
two (three) jets. There are also requirements on the separation between the 
two $b$-jets in the 
$\eta$--$\phi$ space, $\Delta R_{jj}$, to suppress $Z+$jets and $W+$jets
backgrounds as described in Ref.~\cite{Aad:2014xzb}.  
As in the $\ell \ell bb$
channel, the $h$ boson is selected by requiring $105<m_{bb}<145$~GeV.

Additional requirements are imposed on angular quantities sensitive to the
presence of neutrinos in order to suppress the multi-jet
background: the azimuthal angle between 
$\vecmet$ and $\vecmpt$:
$\Delta\phi(\vecmet,\vecmpt)<\pi/2$; 
the minimum azimuthal angle 
between $\vecmet$ and any jet 
${\rm min}[\Delta\phi(\vecmet,{\rm jet})]>1.5$;
and the azimuthal angle between $\met$ and the $b$-jet pair 
$\Delta \phi (\vecmet,bb)>2.8$. The total acceptance times selection
efficiency varies from 4\%, for $m_A = 400$~\gev, to around 7\% for 
the highest $A$ boson masses considered.

It is not possible to accurately reconstruct the invariant mass of the $A$ boson
due to the presence of  neutrinos in the final state. Therefore, 
the transverse mass is used as the final discriminant:
$\mta= \sqrt{(\Et ^{bb} + {\met})^2 - (\vec{p}_{\text{T}}^{\ bb} +
\vecmet)^2}$, where $\Et ^{bb}$ and $\vec{p}_{\text{T}}^{\ bb}$ are 
 the transverse energy and transverse momentum of the $b$-jet pair system. 
As in the $\ell\ell bb$ channel, the resolution is improved 
by scaling each $b$-tagged jet four-momentum by $125~\text{GeV}/m_{bb}$.

%% file: bbbackgrounds.tex
All backgrounds in $\ell\ell bb$/$\nu\nu bb$ final states 
are determined from simulation, apart from the multi-jet background, 
which is determined from data. The multi-jet background in the 
$\mu\mu bb$ final state is
found to be negligible. In the $eebb$ final state, the background is
determined by selecting a sample of events with the electron isolation 
requirement inverted. The sample is normalized by fitting the
$m_{\ell\ell}$ distribution.  In the $\nu\nu bb$ final state, 
the multi-jet background is determined by inverting the 
$\Delta\phi(\vecmet,\vecmpt)$ requirement. The sample is normalized using 
the region with ${\rm min}[\Delta\phi(\vecmet,{\rm jet})]<0.4$.

The $Z$+jets simulated sample is split into different components according
to the true flavour of the jets, i.e. $Z+ll$, $Z+cl$, $Z+cc$, $Z+bl$,
$Z+bc$ and $Z+bb$, where $l$ denotes a light quark ($u$, $d$, $s$) or a gluon.
These components are constrained by defining
control samples which have the same selection as the $\ell\ell bb$ final
state, but with 
the requirements on the number of $b$-tagged jets changed to either zero or
one.
The samples are further divided into events with two or at least three jets.  
In order to improve the description of the data, 
corrections are applied to the simulation as a function of the
azimuthal angle between the two leading jets, $\Delta \phi_{jj}$,   
for $Z+ll$ events and a function of $\ptZ$ for 
the other components, as described in detail in Ref.~\cite{Aad:2014xzb}.

The $W$+jets background, which contributes significantly only in the
$\nu \nu bb$ final state, is split into its components in the same
way as the $Z$+jets sample. It is constrained by defining 
a sample of events that  are selected using the $\met$ triggers 
and contain exactly one lepton
with $\pt>25$~GeV and a tightened isolation requirement.
The transverse momentum of the lepton and ${\vec E}_T^{miss}$
system ($p_T^W$) is required to be above 120 GeV to approximately match
the phase space of the signal region. The sample is split into
events with zero, one or two $b$-tagged jets and 
into events with 2 and 3 jets. A correction depending on $\Delta \phi_{jj}$ is 
applied to $W+ll$ and $W+cl$ events, following studies similar to 
those performed for the $Z$+jets background~\cite{Aad:2014xzb}.

A correction is made to the \pt\ distribution of
$t\bar{t}$ production in the simulation to account for an 
observed discrepancy with the data~\cite{Aad:2014zka}.
The normalization of top-quark pair production in the $\ell\ell bb$ channel 
is measured by defining a sample of events
with exactly one electron and one muon, one of which has $\pt>25$~GeV,
and two $b$-tagged jets with $50 <m_{bb}<180$~GeV.

%% file: bbsummary.tex
\begin{figure*}
  \centering
  \subfigure[$\ell\ell bb$]{
    \includegraphics[width=0.45\textwidth]{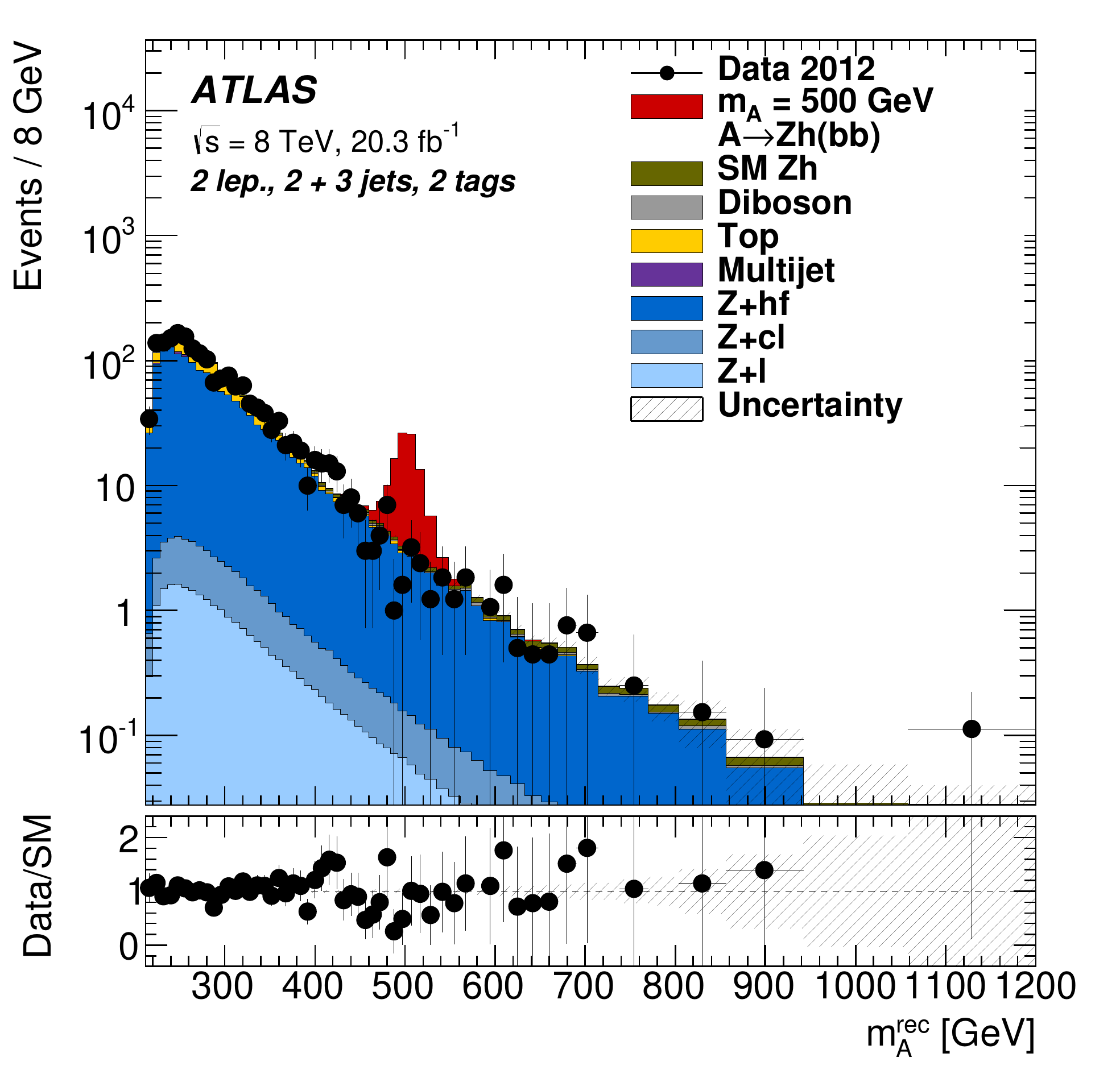}
  }
  \subfigure[$\nu\nu bb$]{
    \includegraphics[width=0.45\textwidth]{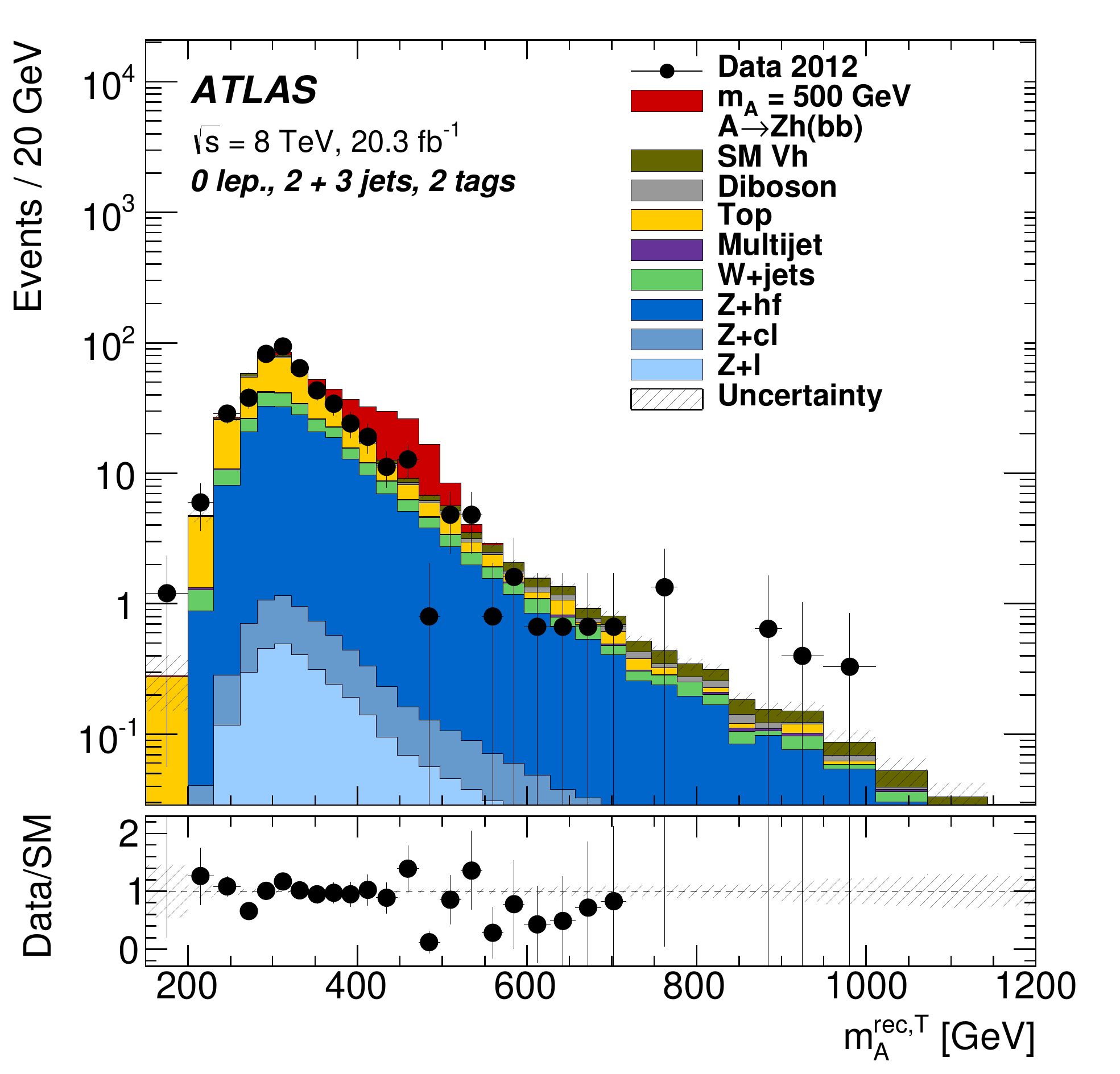}
  }

  \caption{Distributions of the reconstructed $A$ boson mass
    for the $\ell\ell bb$  final state (a)
    and the $A$ boson transverse mass for the $\nu\nu bb$ final state (b). 
    The signal shown in both cases corresponds to
    $\sigma (gg\to A) \times \text{BR}(A\to Zh) \times \text{BR}(h\to bb) = 500$~fb with $m_A$ = 500~GeV.
    The predicted distributions are shown after the profile likelihood 
    fit to the data. The uncertainty is shown as a hatched area, 
    and the overflow is included in the last bin.
}
  \label{fig:xxbb_final}
\end{figure*}

The most important experimental systematic uncertainties 
in the $\ell\ell bb$ and $\nu\nu bb$ final states
come from the jet energy scale uncertainty and the $b$-tagging
efficiency.

The jet energy scale systematic uncertainty 
arises from several sources including
uncertainties from the {\it in situ} calibration, pile-up dependent
corrections and the jet flavour composition~\cite{Aad:2014bia}. 
In addition, an uncertainty on the jet energy
resolution is applied. The jet energy scale and resolution
uncertainties are propagated to the $\met$. The
uncertainty on $\met$ also has a contribution from hadronic energy
that is not associated with jets~\cite{ATLAS-CONF-2013-082}.

The $b$-tagging efficiency uncertainty  depends on jet $\pt$
and comes mainly from the uncertainty on the measurement of the
efficiency in $t\bar{t}$ events~\cite{ATLAS-CONF-2014-004}. 
Similar uncertainties are derived for the
$c$-tagging and light-flavour jet tagging~\cite{ATLAS-CONF-2014-046}.

Other experimental systematic uncertainties that are included but have a smaller
impact are uncertainties from lepton energy scale and 
identification efficiency, the
efficiency of the $\met$ trigger and the uncertainty on the multi-jet
background estimate, which is taken to be 100\% of the estimated
number of events.

In addition to the experimental systematic uncertainties, 
modelling systematic uncertainties are
applied, accounting for possible differences between the data and
the simulation model used for each process. For the
background samples, the procedure described
in Ref.~\cite{Aad:2014xzb} is followed. The $Z$+jets and $W$+jets backgrounds
include uncertainties on the relative fraction of the 
different flavour components,
and on the $m_{bb}$, $\Delta \phi_{jj}$ and $\pt^{Z}$/$\pt^{W}$ distributions. 
For $t\bar{t}$ production, uncertainties  on 
the top-quark transverse momentum, $m_{bb}$, $\met$ and $\pt^{Z}$/$\pt^{W}$ 
distributions are included. 
Uncertainties on the ratio of two-jet to three-jet events 
are also included for each background.

The $m_A^{\text{rec}}$ and $\mta$ distributions for events
passing the $\ell\ell bb$ and $\nu\nu bb$ final-state
selections, respectively, are shown in Fig.~\ref{fig:xxbb_final}.  
The distributions are shown after a profile-likelihood fit, which constrains 
simultaneously the signal yield and the background normalization and shape, 
which is performed in the same manner as in Ref. \cite{Aad:2014xzb}. The overall 
background is more constrained than the individual components, causing the errors 
of indvidual components to be anti-correlated.
The number of events passing the $\ell\ell bb$ and $\nu\nu bb$ final 
state selections are shown in Table~\ref{tab:xxbb_event_yields}, 
where the values for the expectations and 
uncertainties are obtained from the profile-likelihood fit.

\begin{table} 
  \caption{Predicted and observed number of events
    for the $\ell\ell bb$ and $\nu\nu bb$  final states shown
    after the profile likelihood fit to the data. 
  }
  \centering
  \begin{tabular}{lcc}

    \hline
    &  $\ell\ell  bb$   		&  $\nu\nu bb$\\
    \hline

$Z$+jets 		& \phantom{.}1443  $\pm$ 60\phantom{.}  & 225 	$\pm$ 11 	\\
$W$+jets 		&	\phantom{00}--          	& 55  $\pm$ 8 	\\ 
Top 			& \phantom{0.}317  $\pm$ 28\phantom{.}  & 203 $\pm$ 15 	\\							  
Diboson			& \phantom{0}30 $\pm$ 5                 & 10.8  $\pm$ 1.6 	\\ 
SM $Zh$, $Wh$		& \phantom{0}31.7 $\pm$ 1.8             & 22.5  $\pm$ 1.2 	\\ 
Multi-jet		& \phantom{00.}20 $\pm$ 16\phantom{.}   & \phantom{0}3.2   $\pm$ 3.1 	\\ 
Total background	& \phantom{.}1843  $\pm$ 34\phantom{.}  & 521  $\pm$  12 	\\

\hline
Data             &    	1857			&   511           \\
\hline
   \end{tabular}
   \label{tab:xxbb_event_yields}
 \end{table}

%% file: results.tex
\begin{figure*}[ht!]
  \centering
 \subfigure[$A \rightarrow Zh$, $h \rightarrow \tau\tau$]{
    \includegraphics[width=0.45\textwidth]{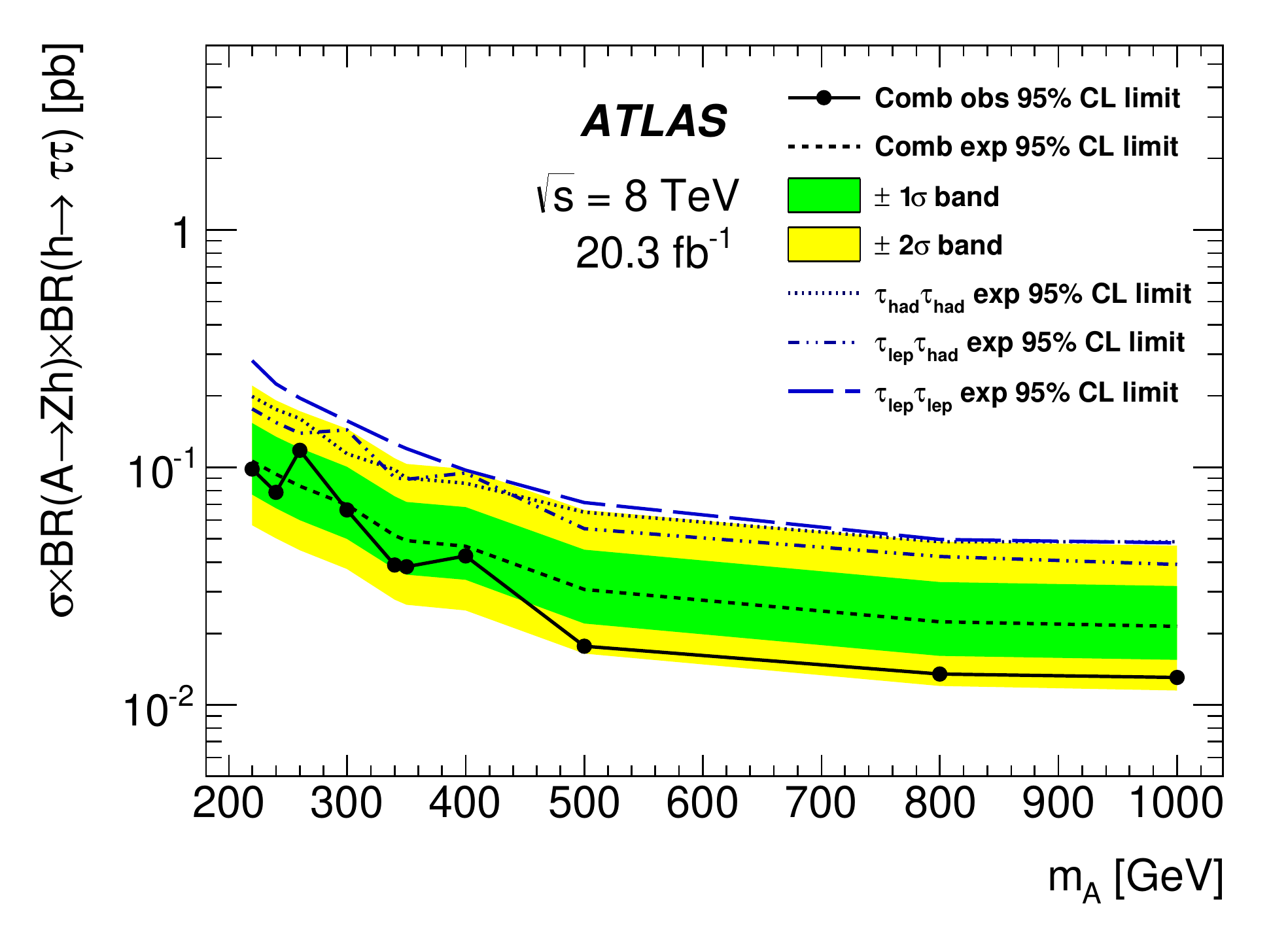}
  }
 \subfigure[$A \rightarrow Zh$, $h \rightarrow bb$]{
    \includegraphics[width=0.45\textwidth]{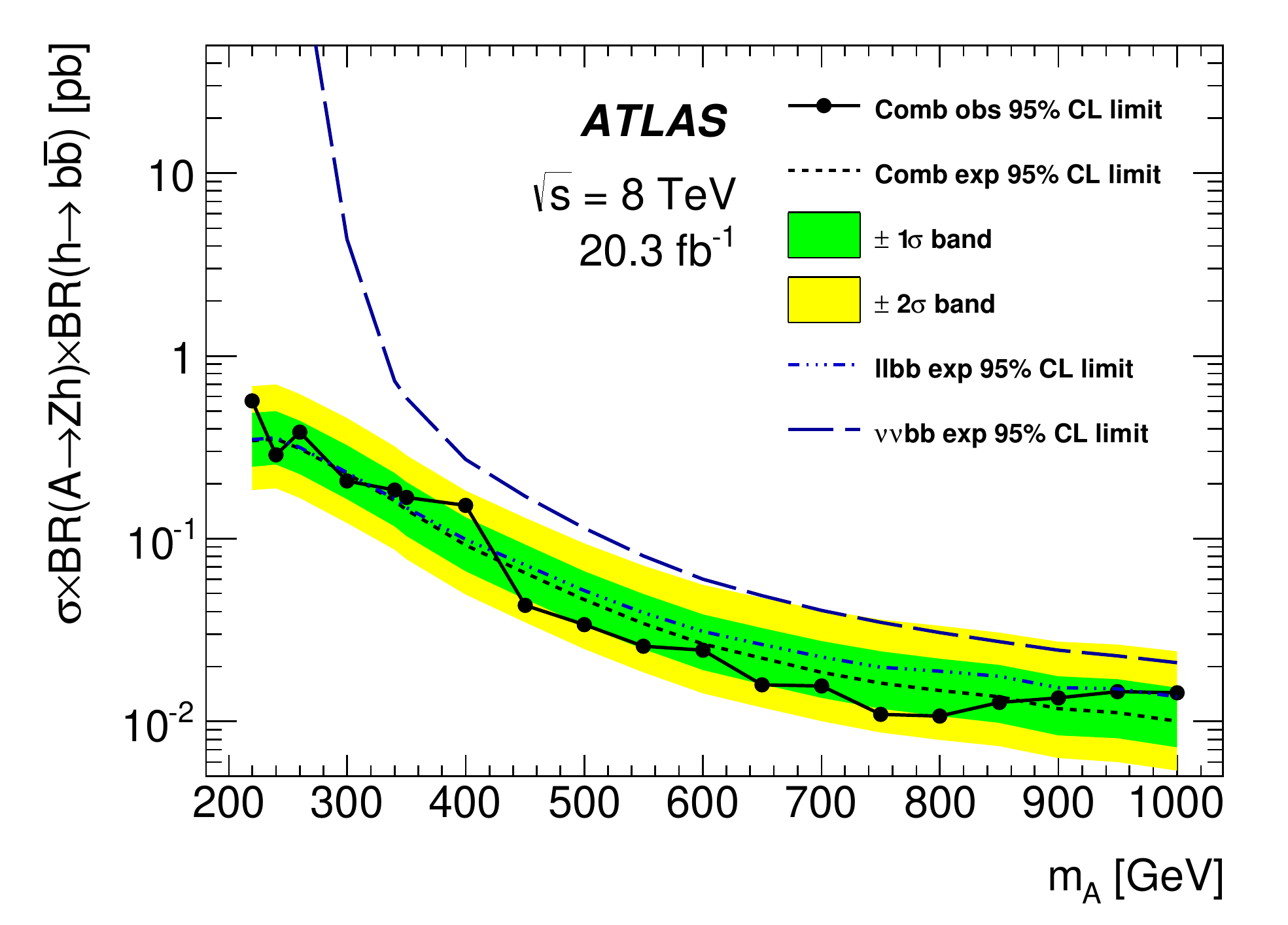}
  }
  \caption{Combined observed and expected upper limits at the 95\% CL for the production
cross section of a gluon-fusion-produced $A$ boson times its
branching ratio to $Zh$ and branching ratio of  $h$ to (a) $\tau\tau$ 
and (b) $bb$. The expected upper limits for subchannels are also shown.}
  \label{fig:limit}
\end{figure*}

\section{Results} 
\label{sec:results} 

\begin{figure*}
  \centering
 \subfigure[Type-I 2HDM, $m_{A} = 300$~GeV]{
    \label{fig:limit-2hdm-1a}
    \includegraphics[width=0.45\textwidth]{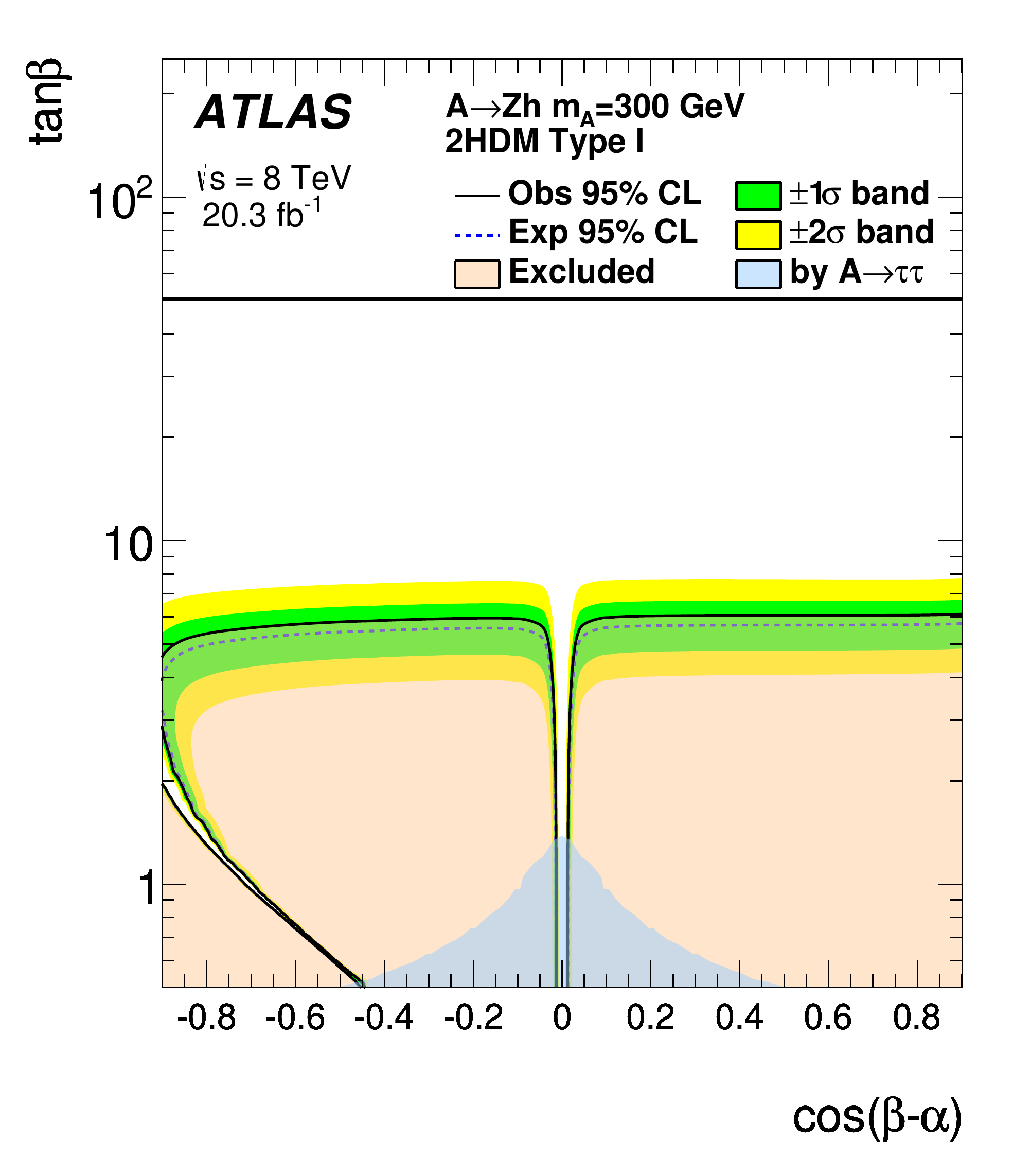}
 }
  \subfigure[Type-II 2HDM, $m_{A} = 300$~GeV]{
    \label{fig:limit-2hdm-1b}
    \includegraphics[width=0.45\textwidth]{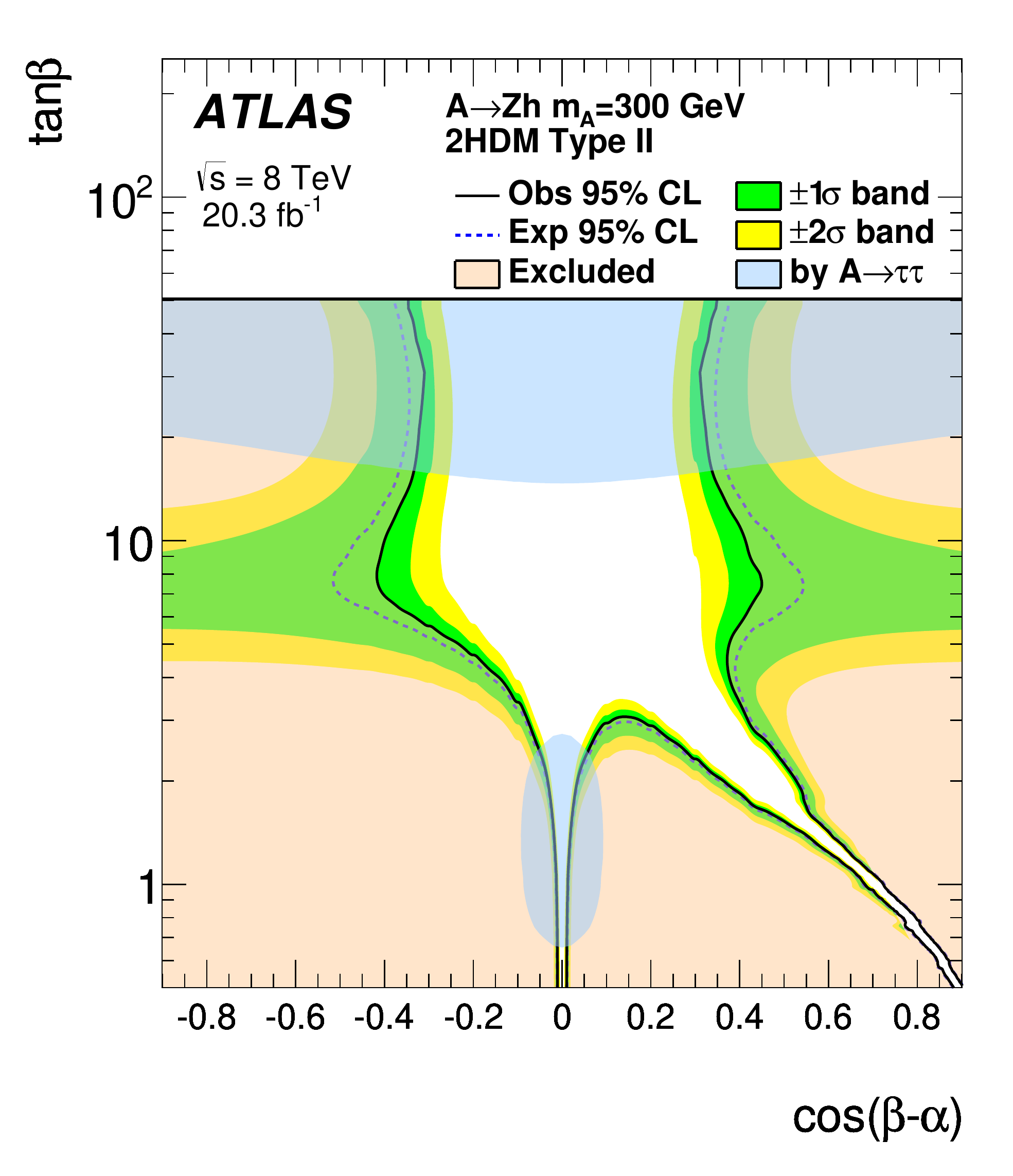}
  }
  \subfigure[Lepton-specific 2HDM, $m_{A} = 300$~GeV]{
    \label{fig:limit-2hdm-1c}
    \includegraphics[width=0.45\textwidth]{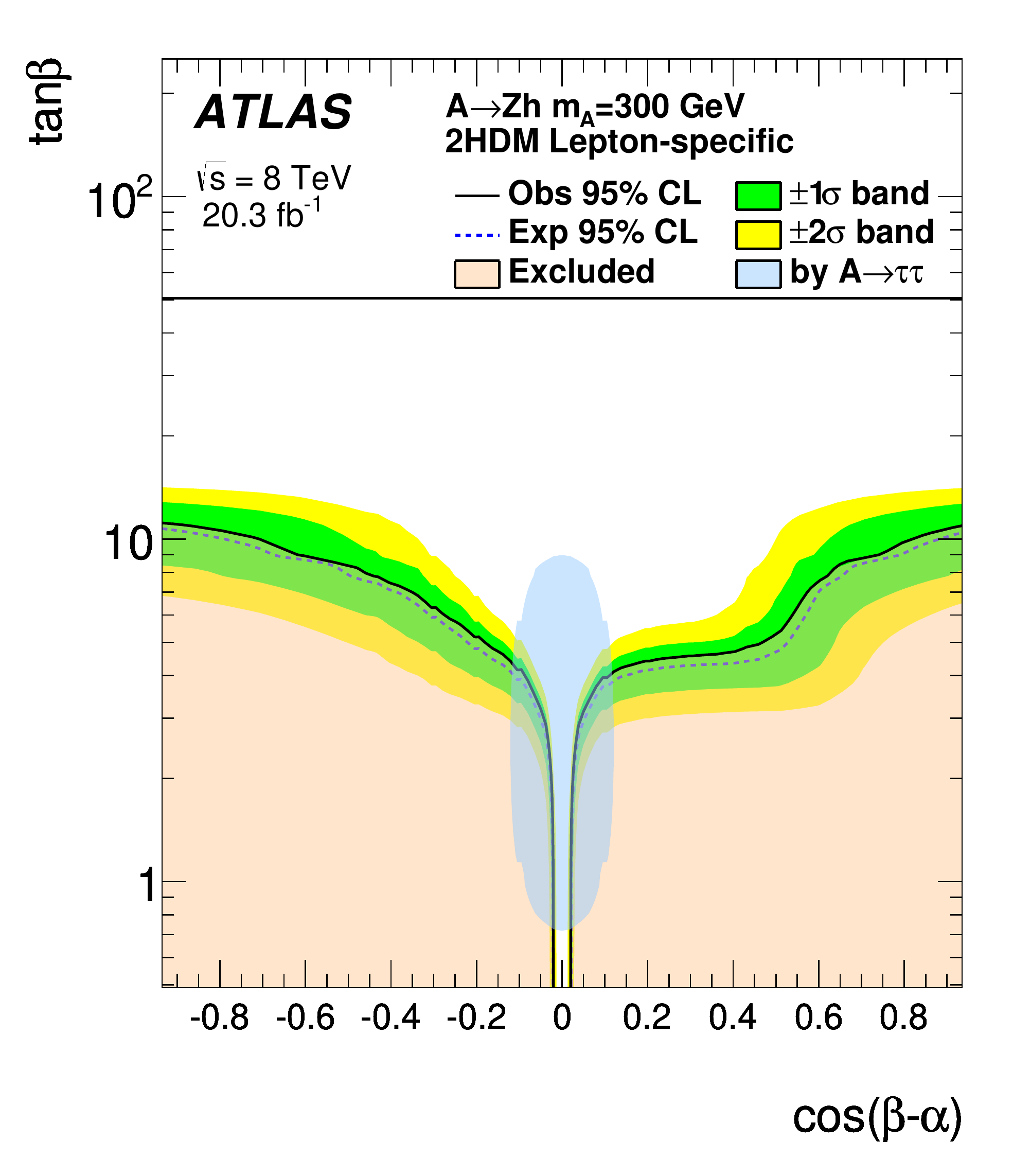}
  }
  \subfigure[Flipped 2HDM, $m_{A} = 300$~GeV]{
    \label{fig:limit-2hdm-1d}
    \includegraphics[width=0.45\textwidth]{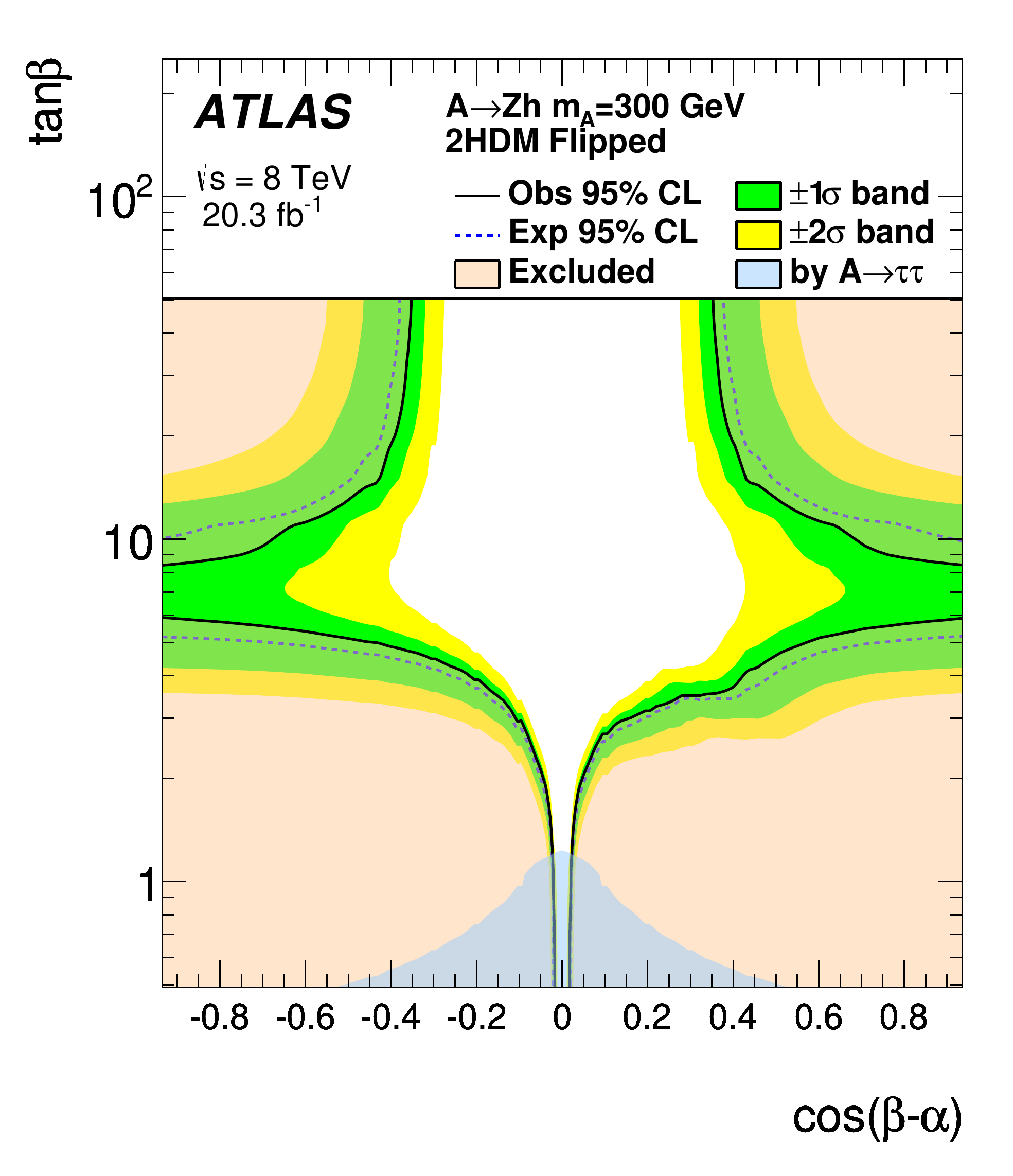}
  }

  \caption{The interpretation of the cross-section limits in the context 
    of the various 2HDM types as a function of the parameters 
    $\tan\beta$ and $\cos(\beta-\alpha)$ for $m_{A} = 300$~GeV: 
    (a) Type-I, (b) Type-II, (c) Lepton-specific, and (d) Flipped.
    Variations of the natural width up to $\Gamma_{A}/m_{A}$=5\% are 
    taken into account.  For  Type-II and Flipped 2HDM, 
    the $b$-associated production is included in addition to 
    the gluon fusion.  The narrow regions with no exclusion power in 
    Type-I and Type-II at low $\tan\beta$ and far from $\cos(\beta-\alpha)=0$ 
    are caused by vanishing branching ratios of
    $h\rightarrow bb$ and/or $h\rightarrow \tau\tau$.
    The blue shaded area denotes the area excluded by taking
    into account the constraints on the CP-odd Higgs boson
    derived by considering the $A\rightarrow \tau\tau$ decay mode
    after reinterpreting the results in Ref.~\cite{Aad:2014vgg}.
}
  \label{fig:limit-2hdm-1}
\end{figure*}

\begin{figure*}
 \centering
  \subfigure[Type-I 2HDM, $\cos(\beta-\alpha) = 0.1$]{
    \label{fig:limit-2hdm-2a}
   \includegraphics[width=0.45\textwidth]{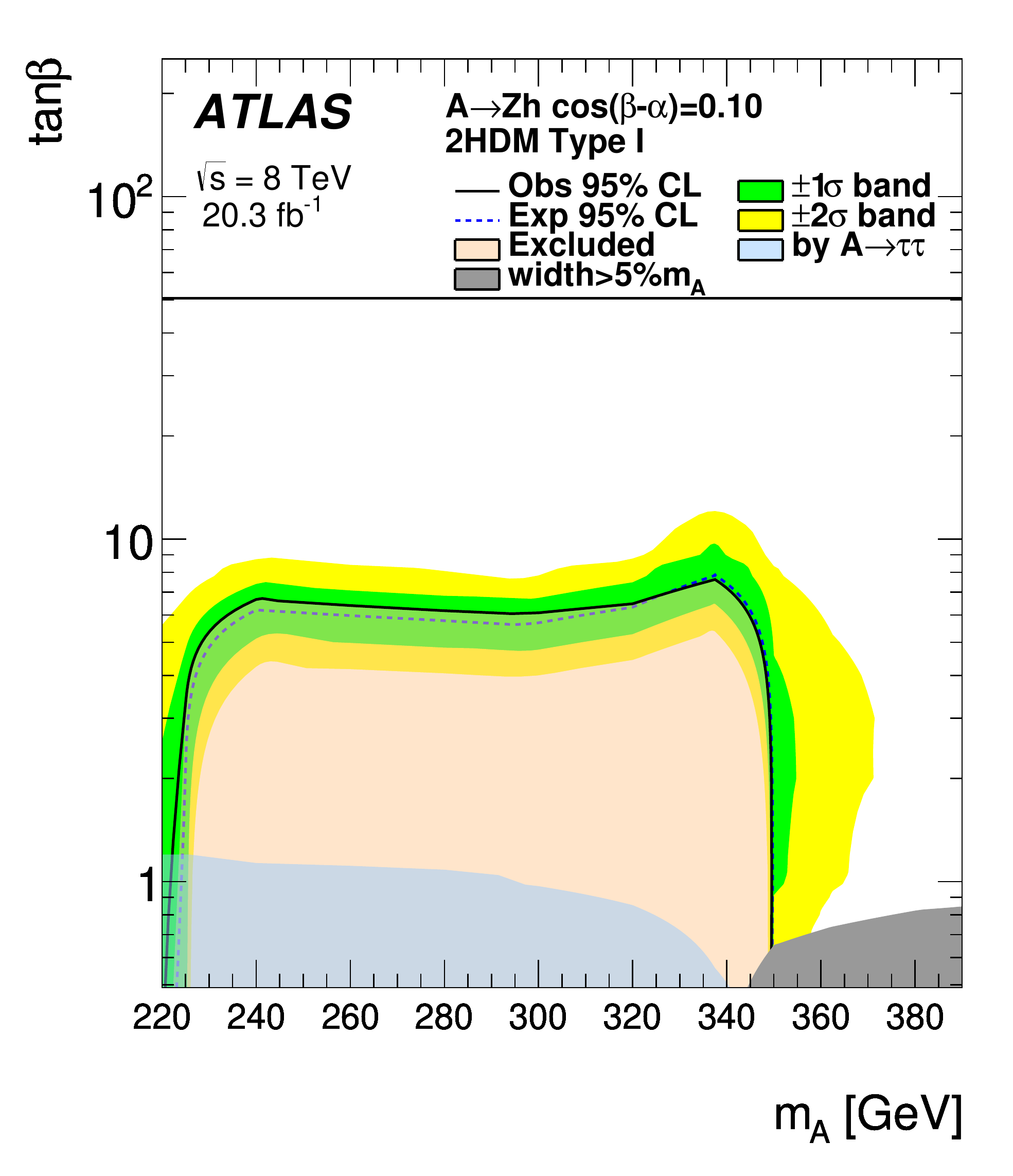}
  }
  \subfigure[Type-II 2HDM, $\cos(\beta-\alpha) = 0.1$]{
    \label{fig:limit-2hdm-2b}
   \includegraphics[width=0.45\textwidth]{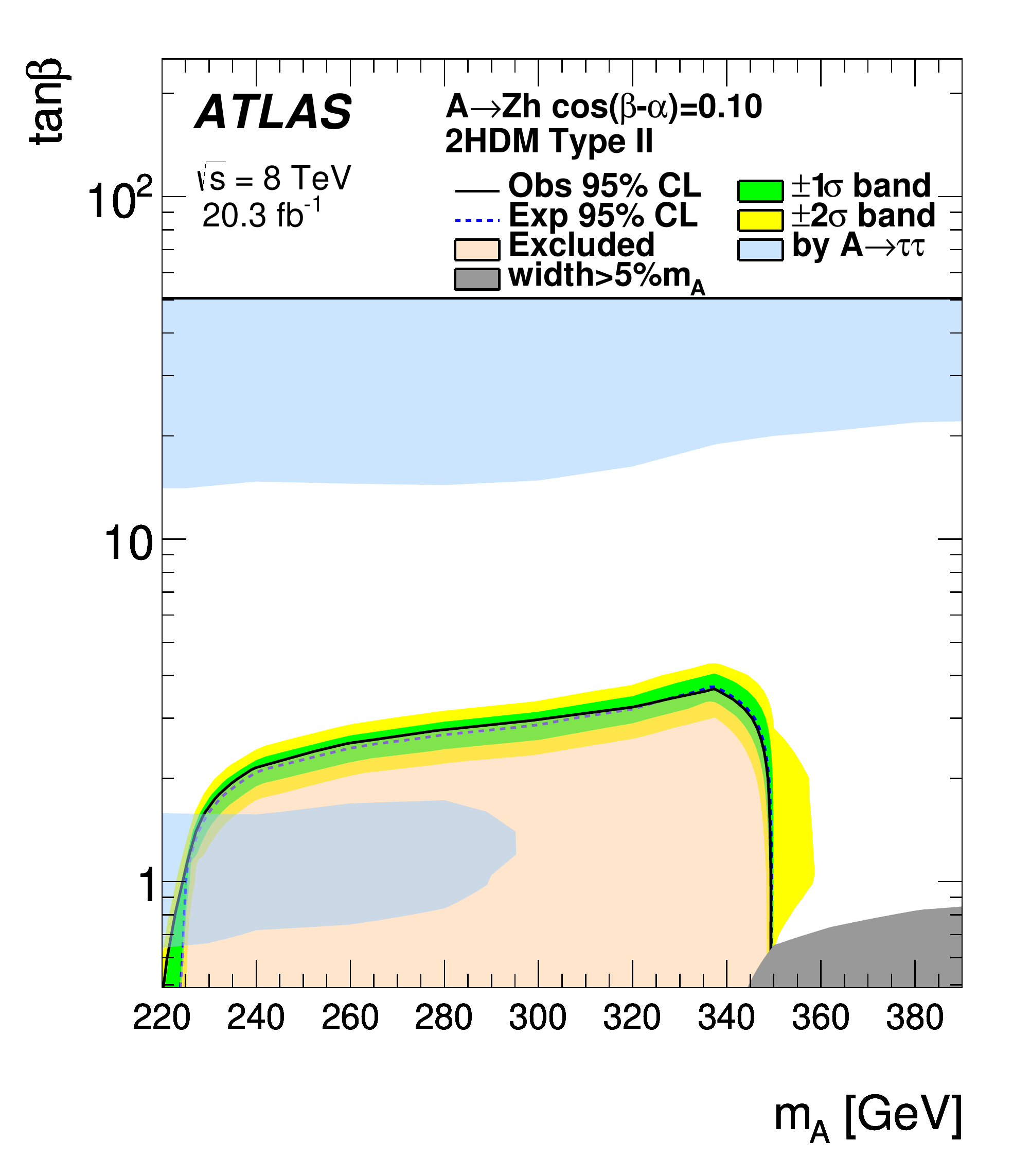}
  }
  \subfigure[Lepton-specific 2HDM, $\cos(\beta-\alpha) = 0.1$]{
    \label{fig:limit-2hdm-2c}
   \includegraphics[width=0.45\textwidth]{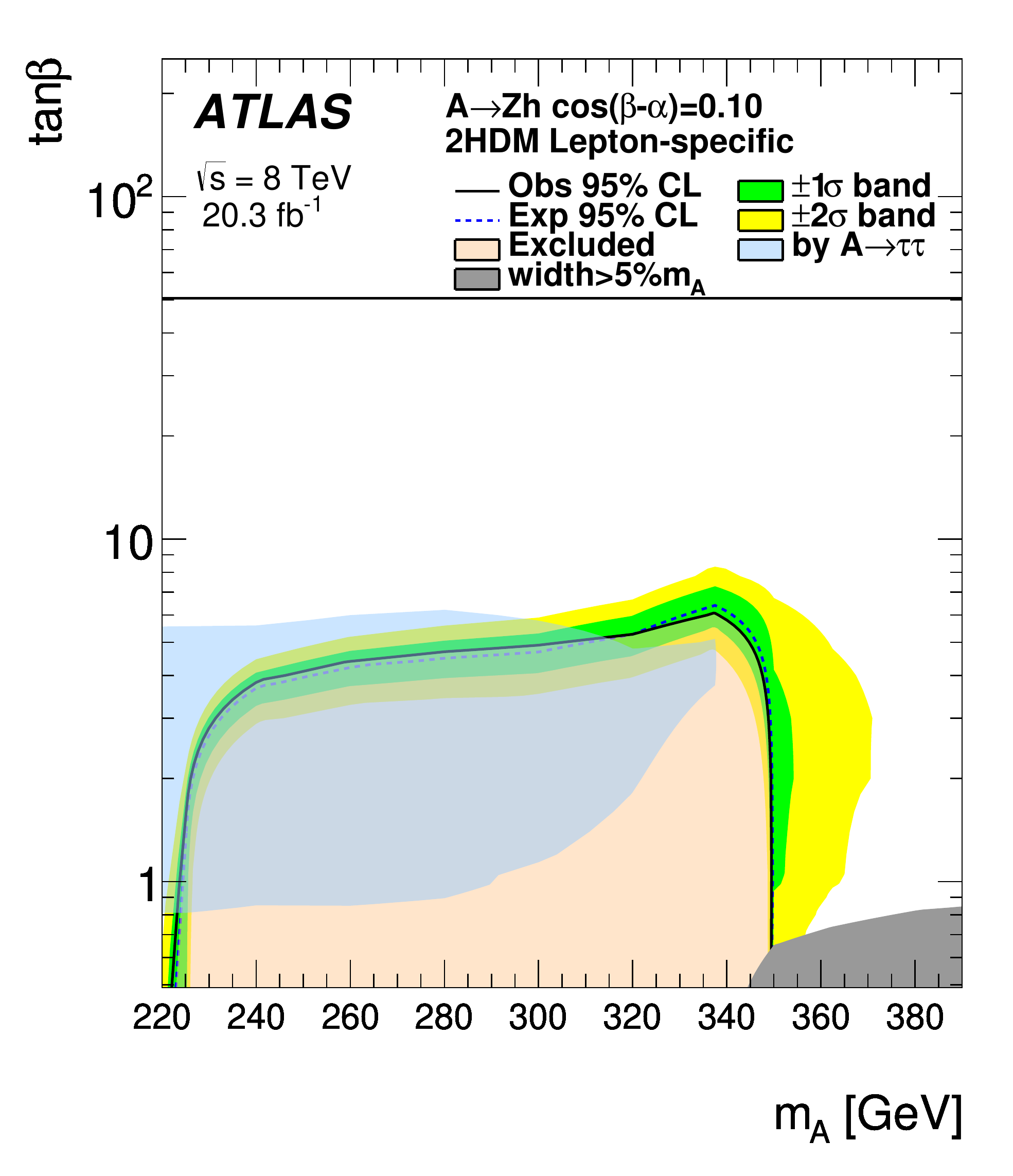}
  }
  \subfigure[Flipped 2HDM, $\cos(\beta-\alpha) = 0.1$]{
    \label{fig:limit-2hdm-2d}
   \includegraphics[width=0.45\textwidth]{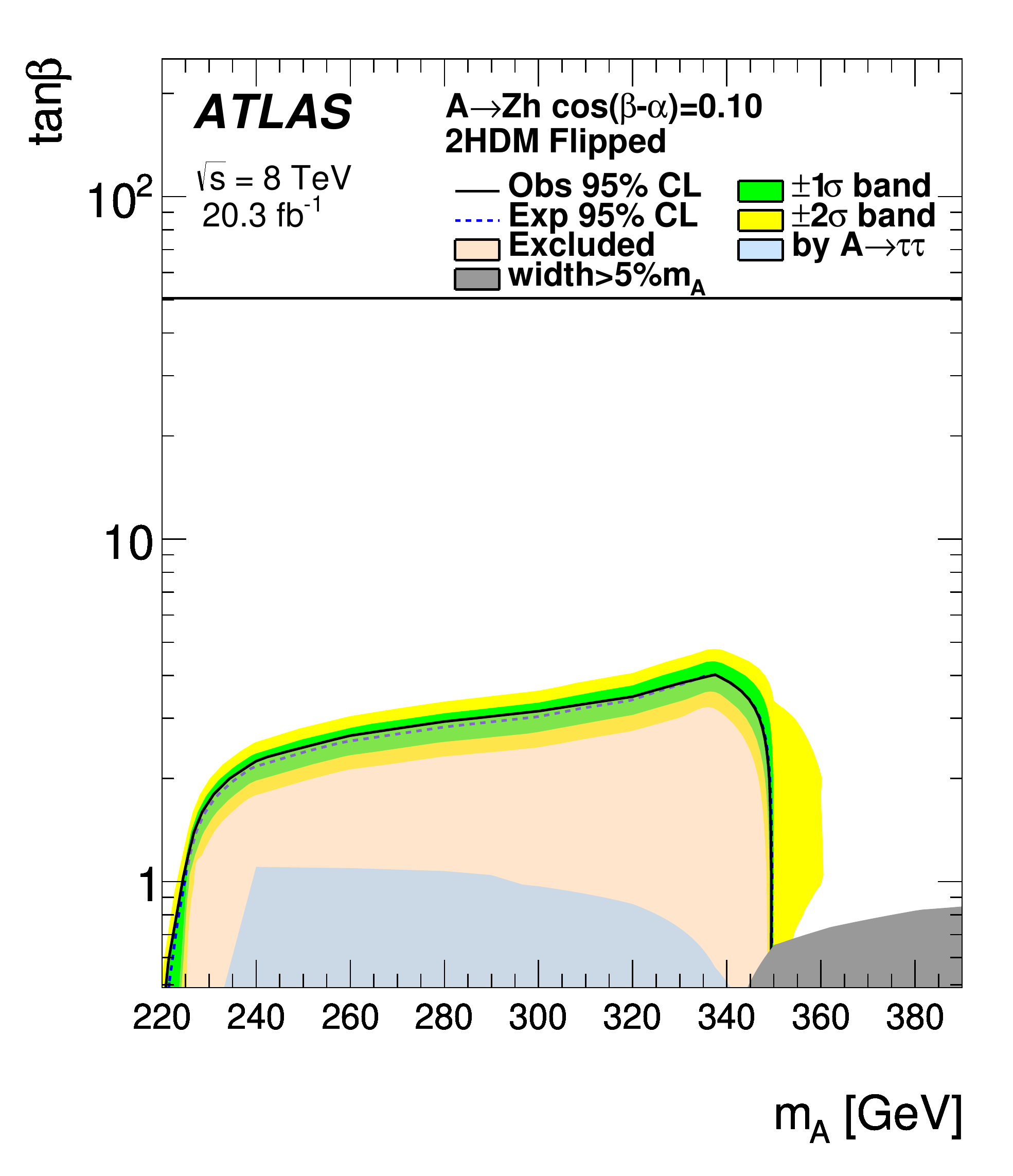}
  }

  \caption{The interpretation of the cross-section limits in the context 
    of the various 2HDM types as a function of the parameters 
    $\tan\beta$ and $m_A$ for $\cos(\beta-\alpha) = 0.1$: 
    (a) Type-I (a), (b) Type-II, (c) Lepton-specific, and (d) Flipped.
    Variations of the natural width up to $\Gamma_{A}/m_{A}$=5\% are 
    taken into account.  The grey solid area indicates that the width is larger than 5\% of $m_{A}$.
    For  Type-II and Flipped 2HDM, 
    the $b$-associated production is included in addition to 
    the gluon fusion.  
    The blue shaded area denotes the area excluded by taking
    into account the constraints on the CP-odd Higgs boson
    derived by considering the $A\rightarrow \tau\tau$ decay mode
    after reinterpreting the results in Ref.~\cite{Aad:2014vgg}.
}
  \label{fig:limit-2hdm-2}
\end{figure*}

In all channels, no significant excess of events  is observed in the data 
compared to the prediction from SM background sources. 
The significance of local excesses is estimated using
$p$-values calculated with a test statistic based on the profile likelihood \cite{Cowan2011}.
The largest data excesses  
are at $m_A = 220$~GeV ($p$-value = 0.014) and $m_A = 260$~GeV 
($p$-value = 0.14) in the combined final states with $h \rightarrow bb$ 
and $h \rightarrow \tau\tau$, respectively. 
Exclusion limits at the 95\% confidence level (CL) are set
on the production cross section times the branching ratio
BR($A\to Zh$) as a function of the $A$ boson mass.
The exclusion limits are calculated with a modified
frequentist method \cite{CLs_2002}, also known as CLs,
and the profile likelihood method, 
using the binned $m_{A}^{\text{rec}}$ 
mass distributions for $\ell\ell\tau\tau$ and $\ell\ell bb$ final states 
and the binned $\mta$ distribution for the  $\nu\nu bb$ final state. 

Fig.~\ref{fig:limit} shows the  95\% CL limits on the production
cross section times the branching ratio, 
$\sigma (gg\to A) \times \text{BR}(A\to Zh)
\times \text{BR}(h\to bb/\tau\tau)$, 
as well as the expected limits for each individual subchannel. 
The limit on the production times the branching ratio is in the range
0.098--0.013~pb and 0.57--0.014~pb for $m_A$ in the range 220--1000~GeV for
the $\tau\tau$ and $bb$ channels, respectively.
The $\tau\tau$ channels use few signal mass points beyond $m_A = 500$~GeV, 
since a coarse binning in $m_{A}^{\text{rec}}$ is adopted in view of the 
very small predicted number of background events.

The results of the search in the $\tau\tau$ and $bb$ channels 
are combined in the context of the CP-conserving 2HDM \cite{Branco:2011iw},
which has seven free parameters and four arrangements of the Yukawa 
couplings to fermions. In particular, the free parameters are the
Higgs boson masses ($m_h$, $m_H$, $m_A$, $m_{H^{\pm}}$),
the ratio of the vacuum expectation values of the two doublets ($\tan\beta$),
the mixing angle between the CP-even Higgs bosons ($\alpha$)
and the potential parameter $m_{12}^2$ that mixes the two Higgs doublets.
The Yukawa coupling arrangements distinguish four different 2HDM models,
determining which of the two doublets,
$\Phi_1$ and $\Phi_2$, couples to up- and down-type quarks and leptons. 
In the Type-I model, $\Phi_2$ couples to all quarks and leptons, whereas in the
Type-II, $\Phi_1$ couples to down-type fermions and $\Phi_2$
couples to up-type fermions. The Lepton-specific model is similar to Type-I
apart from the fact that the leptons couple to $\Phi_1$, instead of
$\Phi_2$. The Flipped model is similar to Type-II apart from 
the leptons coupling to $\Phi_2$, instead of $\Phi_1$. In all these models,
the limit $\cos(\beta-\alpha) \rightarrow 0$ is such that the light CP-even
Higgs boson, $h$, has indistinguishable properties from a SM Higgs boson
with the same mass.
The cross sections for production by gluon fusion are calculated
using {\tt SusHi} \cite{Harlander:2012pb,Harlander:2002wh,Harlander:2003ai,Aglietti:2004nj,Bonciani:2010ms,Harlander:2005rq} 
and the branching ratios  are calculated with 
{\tt 2HDMC} \cite{Eriksson:2009ws}. 
For the branching ratio calculations, it is assumed that
$m_A = m_H = m_{H^{\pm}}$, $m_h = 125$~GeV and 
$m_{12}^2 = m_A^2 \tan\beta/(1+\tan^2\beta)$.

The constraints derived from the combined search in $\tau\tau$ and $bb$ final states are presented
as a function of 2HDM parameters. The exclusion region in the
$\cos(\beta-\alpha)$~versus~$\tan\beta$ plane for $m_A = 300$~GeV 
are shown in Fig.~\ref{fig:limit-2hdm-1} for the four 2HDM models , while
the constraints obtained in the $m_A$--$\tan\beta$ plane  
for $\cos(\beta-\alpha) = 0.10$ are shown in Fig.~\ref{fig:limit-2hdm-2}. 
The width of the $A$ boson in the 2HDM may be larger than the
experimental mass resolution, and it is taken into
account in the 2HDM parameter exclusion regions 
for widths up to 5\% of $m_A$. 
For Type-II and Flipped models, Higgs boson production in association
with $b$-quarks dominates over gluon fusion for large $\tan\beta$ values
($\tan\beta \gtrsim 10$).  The cross section for the $b$-associated
production uses an empirical matching of the cross sections
in the four- and five-flavour  schemes~\cite{Harlander:2011aa}.
Cross sections in the four-flavour scheme are calculated according to
Refs.~\cite{Dawson:2003kb,Dittmaier:2003ej} 
and  cross sections in the five-flavour scheme are calculated using {\tt SusHi}.
The relative efficiencies for the $b$-associated and gluon fusion 
production as well as the predicted cross-section ratio 
are taken into account when  deriving the constraints in 
the two-dimensional planes shown in Fig.~\ref{fig:limit-2hdm-1}. 
The $b$-associated production efficiencies are estimated  
using {\tt PYTHIA8} and {\tt SHERPA} samples.
The regions of parameter space excluded at 95\% CL 
by the $A\rightarrow \tau\tau$ decay mode are displayed in the same plots, using 
the results of a search for a heavy Higgs boson decaying 
into $\tau\tau$ (Ref.~\cite{Aad:2014vgg}), reinterpreted 
considering only the production of an $A$ boson 
via gluon fusion and $b$-associated production.
For $m_A$ values below the $t\bar{t}$ kinematic threshold, 
the search presented here can exclude $\cos(\beta-\alpha)$ values
down to a few percent for $\tan\beta$ values  up to $\approx 3$.

%% file: summary.tex
Data recorded in 2012 by the ATLAS experiment at the LHC, 
corresponding to an integrated luminosity
of 20.3~fb$^{-1}$ of proton--proton collisions at a centre-of-mass energy 8~TeV,
are used to search for a  CP-odd Higgs boson, $A$,
decaying to $Zh$, where $h$ denotes a light CP-even
Higgs boson with a $125\,\gev$ mass.
No deviations from the SM background predictions are
observed in the three final states considered:
 $Zh\to\ell\ell \tau\tau$,  $Zh\to\ell\ell bb$, and  $Zh\to\nu\nu bb$.
Upper limits are set at the 95\% confidence level for 
$\sigma (gg\to A) \times \text{BR}(A\to Zh)
\times \text{BR}(h\to f\bar{f})$ of 0.098--0.013~pb for $f=\tau$ and 
0.57--0.014~pb for $f=b$
in the range of $m_A =$ 220--1000~GeV. 
This  $Zh$ resonance  search 
improves significantly the previously 
published constraints on  CP-odd Higgs boson production
in the low $\tan\beta$ region of the 2HDM.

%% file: Acknowledgements.tex

We thank CERN for the very successful operation of the LHC, as well as the
support staff from our institutions without whom ATLAS could not be
operated efficiently.

We acknowledge the support of ANPCyT, Argentina; YerPhI, Armenia; ARC,
Australia; BMWFW and FWF, Austria; ANAS, Azerbaijan; SSTC, Belarus; CNPq and FAPESP,
Brazil; NSERC, NRC and CFI, Canada; CERN; CONICYT, Chile; CAS, MOST and NSFC,
China; COLCIENCIAS, Colombia; MSMT CR, MPO CR and VSC CR, Czech Republic;
DNRF, DNSRC and Lundbeck Foundation, Denmark; EPLANET, ERC and NSRF, European Union;
IN2P3-CNRS, CEA-DSM/IRFU, France; GNSF, Georgia; BMBF, DFG, HGF, MPG and AvH
Foundation, Germany; GSRT and NSRF, Greece; RGC, Hong Kong SAR, China; ISF, MINERVA, GIF, I-CORE and Benoziyo Center, Israel; INFN, Italy; MEXT and JSPS, Japan; CNRST, Morocco; FOM and NWO, Netherlands; BRF and RCN, Norway; MNiSW and NCN, Poland; GRICES and FCT, Portugal; MNE/IFA, Romania; MES of Russia and ROSATOM, Russian Federation; JINR; MSTD,
Serbia; MSSR, Slovakia; ARRS and MIZ\v{S}, Slovenia; DST/NRF, South Africa;
MINECO, Spain; SRC and Wallenberg Foundation, Sweden; SER, SNSF and Cantons of
Bern and Geneva, Switzerland; NSC, Taiwan; TAEK, Turkey; STFC, the Royal
Society and Leverhulme Trust, United Kingdom; DOE and NSF, United States of
America.

The crucial computing support from all WLCG partners is acknowledged
gratefully, in particular from CERN and the ATLAS Tier-1 facilities at
TRIUMF (Canada), NDGF (Denmark, Norway, Sweden), CC-IN2P3 (France),
KIT/GridKA (Germany), INFN-CNAF (Italy), NL-T1 (Netherlands), PIC (Spain),
ASGC (Taiwan), RAL (UK) and BNL (USA) and in the Tier-2 facilities
worldwide.

%% file: atlas_authlist.tex
\begin{flushleft}
{\Large The ATLAS Collaboration}

\bigskip

G.~Aad$^{\rm 85}$,
B.~Abbott$^{\rm 113}$,
J.~Abdallah$^{\rm 152}$,
S.~Abdel~Khalek$^{\rm 117}$,
O.~Abdinov$^{\rm 11}$,
R.~Aben$^{\rm 107}$,
B.~Abi$^{\rm 114}$,
M.~Abolins$^{\rm 90}$,
O.S.~AbouZeid$^{\rm 159}$,
H.~Abramowicz$^{\rm 154}$
% ...
\bigskip
\\
% ...
$^{11}$ Institute of Physics, Azerbaijan Academy of Sciences, Baku, Azerbaijan\\
% ...
$^{85}$ CPPM, Aix-Marseille Universit{\'e} and CNRS/IN2P3, Marseille, France\\
% ...
$^{90}$ Department of Physics and Astronomy, Michigan State University, East Lansing MI, United States of America\\
% ...
$^{107}$ Nikhef National Institute for Subatomic Physics and University of Amsterdam, Amsterdam, Netherlands\\
% ...
$^{113}$ Homer L. Dodge Department of Physics and Astronomy, University of Oklahoma, Norman OK, United States of America\\
$^{114}$ Department of Physics, Oklahoma State University, Stillwater OK, United States of America\\
% ...
$^{117}$ LAL, Universit{\'e} Paris-Sud and CNRS/IN2P3, Orsay, France\\
% ...
$^{152}$ Institute of Physics, Academia Sinica, Taipei, Taiwan\\
% ...
$^{154}$ Raymond and Beverly Sackler School of Physics and Astronomy, Tel Aviv University, Tel Aviv, Israel\\
% ...
$^{159}$ Department of Physics, University of Toronto, Toronto ON, Canada\\
% ...
\end{flushleft}